\documentclass{IEEEtran}
\usepackage{cite}

\usepackage{graphicx}
\DeclareGraphicsExtensions{.eps,.pdf}

\usepackage{tikz}
\usetikzlibrary{arrows}

\usepackage{balance}
\usepackage{amsmath}
\usepackage{amsthm}
\usepackage{amsfonts}
\usepackage{amssymb}
\usepackage{algorithm,tabularx}
\usepackage{algpseudocode}

\makeatletter
\newcommand{\multiline}[1]{%
  \begin{tabularx}{\dimexpr\linewidth-\ALG@thistlm}[t]{@{}X@{}}
    #1
  \end{tabularx}
}
\makeatother

\usepackage{bm}
\usepackage{soul}
\usepackage{color}
\usepackage{optidef}
\usepackage{enumitem}
\newtheorem{theorem}{Theorem}

\newcommand{\RNum}[1]{\uppercase\expandafter{\romannumeral #1\relax}}

\makeatletter
\newcommand{\removelatexerror}{\let\@latex@error\@gobble}
\makeatother

\begin{document}
%
\title{Efficient, Fair and QoS-Aware Policies for Wirelessly Powered Communication Networks}



\author{Roohollah~Rezaei,~\IEEEmembership{Student Member,~IEEE,} Naeimeh Omidvar,~\IEEEmembership{Member,~IEEE,} Mohammad Movahednasab,  Mohammad~Reza~Pakravan,~\IEEEmembership{Member,~IEEE,} Sumei~Sun,~\IEEEmembership{Fellow,~IEEE,}
       and Yong~Liang~Guan,~\IEEEmembership{Senior Member,~IEEE.}
\thanks{ Part of this work has been presented in IEEE Global Communications Conference (GLOBECOM) 2018 \cite{Rezaei2018OptandNearOpt}. The first author acknowledges the support of the Agency for Science, Technology and Research (A*STAR) Research Attachment Program (ARAP).  
R. Rezaei and S. Sun are with the Institute for Infocomm Research, Singapore
138632 (e-mail: sturr@i2r.a-star.edu.sg; sunsm@i2r.a-star.edu.sg).
R. Rezaei, M. Movahednasab, N. Omidvar, and M. R. Pakravan are with the Electrical Engineering Department, Sharif University of Technology, Tehran, Iran (e-mail: rezaei\_roohollah@ee.sharif.edu; movahednasab@ee.sharif.edu; n\_omidvar@ee.sharif.edu; pakravan@sharif.edu). 
Y. Guan is with the School of Electrical and Electronic
Engineering, Nanyang Technological University, Singapore (e-mail: eylguan@ntu.edu.sg).
}}        

\maketitle

\begin{abstract}
Wireless {\color{black}power transfer (WPT)} is a viable source of energy for {\color{black}wirelessly powered communication networks (WPCNs).} In this paper, we first consider WPT from an {\color{black}energy access point (E-AP) to multiple energy receivers (E-Rs)} to obtain the optimal policy that maximizes the WPT efficiency. For this purpose, we formulate the problem of maximizing the total average received power of the E-Rs subject to the average and peak power level constraints of the E-AP. The formulated problem is a non-convex stochastic optimization problem. Using some stochastic optimization techniques, we tackle the challenges of this problem and derive a closed-form expression for the optimal solution, which requires the explicit knowledge of the distribution of {\color{black}channel state information (CSI)} in the network. We then propose a near-optimal algorithm that does not require any explicit knowledge of the CSI distribution and prove that the proposed algorithm attains a near-optimal solution within a guaranteed gap to the optimal solution. 
{\color{black}We next} consider fairness among the E-Rs and propose a {\color{black}quality of service} (QoS) aware fair policy that maximizes a generic network utility function while guaranteeing the required QoS of each E-R. Finally, we study a practical {\color{black}wirelessly} powered communication scenario in which the E-Rs utilize their energy harvested through WPT to transmit information to the E-AP. We optimize the received information at the E-AP under its average and peak transmission power constraints and the fairness constraints of the E-Rs. Numerical results show the significant performance of our proposed solutions compared to the state-of-the-art baselines.
\end{abstract}



%
\IEEEpeerreviewmaketitle

\section{Introduction}

 
%
%
%
%
%
%
%

Providing energy resources for wireless devices 
is a critical issue
in many emerging applications. For example, in sensor networks, recharging the batteries of  wireless nodes is a costly and time-consuming process. In some other applications such as medical implants inside human bodies, {replacement of the batteries} is highly difficult and almost impractical. To {\color{black}address} the issues mentioned above, wireless power transfer (WPT) is proposed as a key enabling technology to provide continuous, stable, and controllable energy {resources} to wireless devices over the air \cite{Kang2015}. 
This technology is currently being incorporated in many devices and applications, such as mobile phones,  electric toothbrushes, wirelessly powered drones, wireless charging stations for {\color{black}electric} vehicles, and {\color{black}wirelessly} transfer of the power gathered by solar-panel arrays in the space, to name just a few \cite{siddabattula2015not}. 

WPT can be used as the source of energy for wirelessly powered communication networks (WPCNs). In {\color{black}WPCNs, an energy access point (E-AP)} transfers
energy to a number of wireless nodes. Then, the nodes can {\color{black}utilize} the harvested energy to transmit their
information back to the {\color{black}E-AP} \cite{zhong2018energy, chi2017minimization , Rezaei2018PIMRC, chu2017wireless, Rezaei2019Secrecy}. The energy is transferred by magnetic induction or radio frequency (RF). The latter technique covers longer transmission ranges, requires a simpler {\color{black}structure for the receivers,} and also better supports multiple receivers than the former technique \cite{Mou2015}. Considering these advantages and {\color{black}following} many previous works (e.g., see \cite{Yang2014, Lee2017, Xu2014, Zeng2015, rezaei2018optimal, Movahednasab2019}), throughout this paper, we focus on RF-based WPT (RF-WPT).

{In order to improve the performance of WPCNs, optimization of WPT policy is crucial.} 
To optimize the WPT policy, the E-AP needs to know 
the time-varying channel state information (CSI) of its outgoing links, which is a random variable with usually unknown distribution, in practice. Most of the existing works in the literature consider the CSI of {\color{black}the network} as a known deterministic parameter in each timeslot. As a consequence of such {\color{black}a} naive simplification, they can formulate the problem of finding the optimal WPT policy {\color{black}with} a deterministic optimization problem (e.g., see \cite{Yang2014,Bi2015}), which is then solved in each timeslot, independently.    
However, such short-term solutions lack a global view of the long-term {CSI} and fail to incorporate the long-term channel fluctuations in the optimization of the transmission policy. For example, consider the case when a channel's condition is poor in a particular timeslot. Under a short-term optimization policy, the transmission resources cannot be preserved for a more effective {utilization} in the upcoming timeslots {\color{black}that} may have better CSI. In contrast, the long-term solutions, obtained via the long-term optimization of the policy, can avoid transmission in the case of poor channel {\color{black}conditions and} save the energy to be used for transmission in the later timeslots when the CSI is better.

Despite the aforementioned advantages of the long-term WPT solutions, there are still very few works on long-term optimization in the related literature \cite{Choi2018,aboelwafa2019towards,lyu2017backscatter,Li2019,Biason2016,Choi2015}. In \cite{Choi2018}, the authors have considered an E-AP that transmits energy toward sensor nodes. The E-AP retains the average energy of the sensor nodes near a constant value to keep them alive. The works in \cite{aboelwafa2019towards} and \cite{lyu2017backscatter} have studied the {\color{black}optimization of the average throughput} in a finite number of timeslots. The authors in \cite{Li2019} 
have considered an E-AP that transfers energy to one single-antenna node. The node stores {\color{black}and then utilizes the received energy  to transmit its information toward the E-AP.} 
 Biason et al. in \cite{Biason2016} have studied an E-AP that transfers energy to two {nodes and receives} their uplink information. Using Markov decision theory, they have proposed a transmission policy for the E-AP that maximizes the minimum received information rate of the nodes. Yet, none of the above works have considered an infinite time horizon optimization problem for more than two nodes.   
Finally, Choi et al. in \cite{Choi2015} have investigated a scenario in which the E-AP aims at stabilizing the data queues of several battery-operated single-antenna nodes while consuming the minimum transmission power.

In this paper, we first {\color{black}focus on improving} WPT efficiency {\color{black}in the network} by maximizing the average total received power of the E-Rs subject to the maximum and average power budget of the E-AP. To this end, we first propose a stochastic optimization formulation for
 the long-term optimization of WPT policy. 
The stochastic optimization formulation is non-convex {\color{black}and hence, highly} non-trivial to solve. To address the challenges of the formulated problem, we use some stochastic optimization techniques and propose the optimal and near-optimal solutions for the formulated problem. 
Furthermore, as the power budget of the E-AP is limited, maximizing the total received power of the E-Rs may lead to severe unfairness among them, due to the near-far problem \cite{Bi2015}. As {\color{black}such, in} order to maintain the fairness among the E-Rs, we next propose a quality of service (QoS) aware fair WPT policy that maximizes a generic network utility function while guaranteeing the required QoS of each E-R. The considered network utility function includes many {well-known fairness models, such as max-min fairness, proportional fairness, and $\alpha$-fairness \cite{Omidvar2018Optimal}.}
{Finally,} we {\color{black}focus on} WPT efficiency in a generic WPCN, where the E-AP first transmits power to all the E-Rs, and then the E-Rs {\color{black}utilize their harvested energy} to transmit information in the uplink to the E-AP. The average received information is maximized subject to the fairness constraints and the E-AP's average and peak power constraints. 
Compared to the previous works, we propose an algorithm that does not need the distribution of the CSI, includes several well-known fairness criteria, {\color{black}and can be applied to a wider range of scenarios, i.e., with any number of multi-antenna E-Rs.} It also considers maximizing the received information over an infinite number of timeslots. 

Note that the considered scenarios in this paper can be deployed in many {\color{black}practical applications in the area of Internet of Things (IoT)}, such as smart home and smart factory. The sensor nodes in these applications are placed in different places to monitor the {\color{black}environmental} conditions, such as air pressure, temperature, {\color{black}and humidity} \cite{zhong2018energy,shaviv2016capacity}. These nodes are energy-constrained and rely on the harvested energy from a central node to continue their operation, as shown in an experimental environment in \cite{Choi2018}. They may also transmit their information to the central node for monitoring purposes, for instance, via a web portal as implemented in \cite{gunawan2017prototype}.

The main contributions of this paper can be summarized as follows:

\begin{itemize}

\item A novel stochastic optimization problem formulation is proposed for WPT of the E-AP to the E-Rs, which aims at optimizing the long-term performance of the WPT efficiency.


\item 
A closed-form expression for the optimal WPT policy is derived. In addition, a near-optimal algorithm is also proposed, which does not require the CSI distribution. {\color{black}Moreover,} the optimality gap of the proposed algorithm is analytically derived, which can be made arbitrarily small.



\item Furthermore, to ensure fairness among the E-Rs, a generic fair WPT problem formulation  is {\color{black}considered, and} a near-optimal power transfer policy is proposed for the formulated stochastic optimization problem.
\item
Finally, a generic {\color{black}wirelessly} powered communication scenario is studied, where the harvested energy by the E-Rs is then {\color{black}utilized} to transmit their data in the uplink to the E-AP.  The proposed scenario considers fairness constraints, guarantees a minimum average throughput for each E-R, and {\color{black}dynamically adjusts the portions of each timeslot that are going to be used by each E-R for its information transfer or  energy harvesting.} We propose a near-optimal power transfer and time allocation policy to maximize the total throughput of all the E-Rs.
\end{itemize}


{\color{black}The paper is organized as follows:}  Section \RNum{2} introduces the system model. The
proposed stochastic optimization problem formulation for maximizing the long-term power transfer efficiency as well as the proposed optimal and near-optimal solutions  are presented in Section \RNum{3}. 
Section \RNum{4} considers fairness among the E-Rs and presents the proposed fair near-optimal solution. Section \RNum{5} illustrates the considered generic {\color{black}wirelessly} powered communication scenario and presents the associated formulated problem and the proposed solution.
Numerical results are presented in Section \RNum{6}. Finally, Section \RNum{7} concludes the paper.

\section{System Model} \label{sec:SysMod}
\begin{figure}
\centering
\includegraphics[width = 0.9\linewidth]{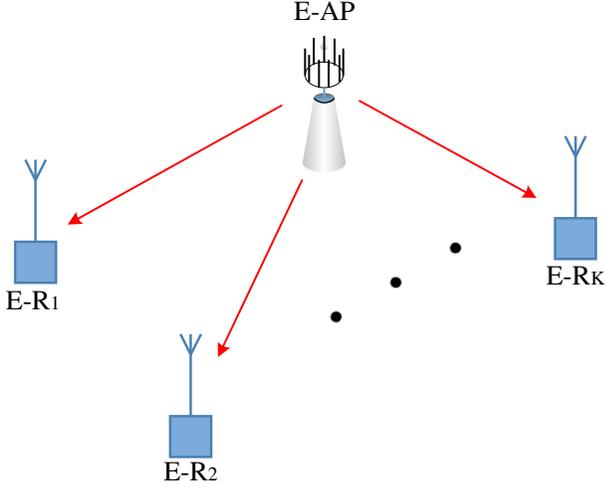}
\caption{Illustration of the considered system model.}
\label{fig:SysMod}
\end{figure}

We consider a network consisting of one E-AP (that is connected to a stable power source) and $K$ E-Rs, as shown in Fig. \ref{fig:SysMod}. The E-AP and the E-Rs  are  equipped with $N$  and $M$ antennas, respectively, where $N >M$.
 The E-AP transfers energy to the E-Rs by transmitting a tone signal (for the sake of saving bandwidth) and {\color{black}employing beamforming techniques (e.g., see \cite{chu2017wireless, Rezaei2019Secrecy, Yang2014})} to  focus the transmitted power  toward the  E-Rs. 
\begin{figure}
\centering
\includegraphics[width = 0.9\linewidth]{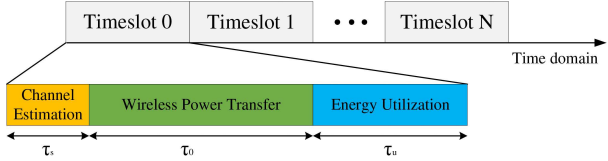}
\caption{The considered time-slotted structure for the time domain.\vspace{-10 pt}}
\label{fig:timeslotView}
\end{figure}

As illustrated in Fig. \ref{fig:timeslotView}, we consider a time-slotted system in which the  time domain is divided into timeslots of fixed length. 
{\color{black}At} the beginning of each timeslot, a small portion of the timeslot (with a fixed duration of $\tau_s$) is reserved for channel estimation of the outgoing channels by the E-AP. The CSI can be estimated via pilot aided methods {\color{black}(e.g., see \cite{Yang2014}).} The rest of the timeslot is divided into two phases used for wireless power transfer from the E-AP to the E-Rs (WPT phase) and sequential information transfer from the E-Rs to the E-AP (energy utilization phase), respectively. The duration of each of the two phases (denoted by $\tau_0$ and $\tau_u$, respectively) is dynamically changing according to the WPT policy in order to maximize the WPT efficiency in the network. 
Note that since the channel
estimation is beyond the scope of this paper, we assume $\tau_s$ to be fixed. Moreover, without loss of generality, we assume that $\tau_0 + \tau_u = 1$. Finally, similar to the previous works  (e.g., see\cite{Yang2014, Ju2014c}), we consider a quasi-static flat-fading channel model for the channels between the E-AP and the E-Rs,  
where the CSI remains constant during each timeslot and varies from one timeslot to the next one.

\subsection{Power Transfer by the E-AP}   
In the WPT phase, of each timeslot $l$, the
 transmitted signal from the E-AP, denoted by $\bm{x} \in \mathbb{C}^{N\times1}$, is determined by the adopted WPT policy of the E-AP. 
The received signal of $ \text{E-R}_i $ in timeslot $ l $ is given by 
\begin{align}
\bm{y}_i(t) &= \bm{H}_i[l]\bm{x}[l]+\bm{z}_i(t),\ \ lT+\tau_{s} \le t < lT+\tau_{s}+\tau_0, \nonumber \\
&\forall i \in \{ 1,...,K \},
\end{align} 
where $T$ is the length of a timeslot, $\bm{y}_i\in \mathbb{C}^{M\times1}$ denotes the baseband signal of E-R$_{i}$, $ \bm{x}[l] $ is the signal of the E-AP in timeslot $ l $, and $\bm{z}_i \in \mathbb{C}^{M\times 1}$ represents the noise at E-R$_{i}$. Moreover, $\bm{H}_i[l]$ denotes the equivalent baseband channel {\color{black}matrix of the links} between the E-AP and $ \text{E-R}_i $ in the $l^{th}$ timeslot. 
It is a complex matrix where its $(m,n)$ entry represents the CSI 
 of the link between the $m^{th}$ antenna of $ \text{E-R}_i $ and the  $n^{th}$ antenna of the E-AP. This channel matrix remains constant during a timeslot and is independent and identically distributed (i.i.d.) in successive timeslots. {\color{black}Moreover, $\bm{H}[l] \triangleq (\bm{H}_1[l],...,\bm{H}_K[l])$ represents the CSI of the network in timeslot $l$, and
 $\bm{H}^{(l)} \triangleq (\bm{H}[0],\bm{H}[1],...,\bm{H}[l])$ represents the CSI history of the network until timeslot $l$.}

\subsection{The Energy Reception by the E-Rs}
\begin{figure}
\centering
\includegraphics[width = 0.7\linewidth]{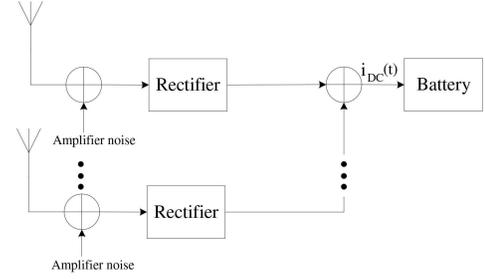}
\caption{The structure of an E-R.\vspace{-10 pt}}
\label{fig:RecvStr}
\end{figure}

The structure of an E-R is shown in Fig. \ref{fig:RecvStr}. As {can be} seen in this figure, the E-R first utilizes a rectifier to convert the received RF signal to a DC current. This current then charges the battery of the E-R.  The amount of the harvested energy of $\text{E-R}_i$ during timeslot $ l $ is denoted by $Q_i[l] $. Note that similar to the previous works (e.g., see \cite{Bi2014}), we neglect the energy contribution of noise. Therefore, $Q_i[l]$ can be written as follows:
\vspace{-5 pt}
\begin{align}
Q_i[l] =  \eta \tau_0 \left\lVert \bm{H}_i[l]\bm{x}[l]  \right \rVert^2 &= \eta \tau_0 Tr(\bm{W}_i[l]\bm{x}[l]\bm{x}^H[l]), \nonumber \\ &\forall i = 1,2,...,K,   \label{equ:receivedEnergyEq2}
\end{align}
where $\bm{W}_i[l]\triangleq\bm{H}_i^{H}[l] \bm{H}_i[l]$ and $\eta \in [0,1)$ represents the energy conversion efficiency. Moreover, $Tr(\bm{A})$ and $\bm{A}^H$ are the trace and transpose hermitian of square matrix $\bm{A}$, respectively.

\subsection{\color{black}Energy Utilization for the E-Rs' Information Transfer}
\begin{figure}
\centering
\includegraphics[width = 0.9\linewidth]{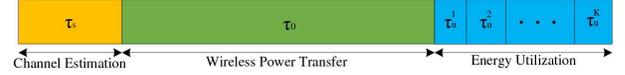}
\caption{The considered structure for each timeslot.\vspace{-10 pt}}
\label{fig:WPCNTimeSlot}
\end{figure}

In Section \ref{sec:WPCN}, we consider a generic {\color{black}wirelessly} powered communication scenario 
that in each timeslot,  the E-Rs utilize their harvested energy during the WPT phase  to transmit their information sequentially to the E-AP, during the energy utilization phase.
As shown in Fig. \ref{fig:WPCNTimeSlot}, the energy utilization phase duration ($\tau_u$) is shared {\color{black}by} the E-Rs for their information transfer to the E-AP, in a sequential way. Each E-R$_i$ is allocated with a duration of $\tau_u^i$ which is dynamically determined by the joint energy and information {\color{black}transmission} policy. Consequently, the throughput of E-R$_{i}$ in timeslot $l$ is    
%

\begin{equation}
D_i[l] = \tau^i_u[l] \log |\bm{I} + \bm{H}'_i[l] \bm{S}_i[l]\bm{H}_i^{'H}[l]|,\ \forall i = 1,2,...,K, \label{equ:InfoOneSlot}
\end{equation}  
where $\bm{S}_i[l]$ is the covariance matrix of the transmission signal of E-R$_i$, and $\bm{H}'_i[l]$ is the uplink {\color{black}channel matrix of E-R$_i$.} 



\section{\color{black}Problem Formulation and the Proposed WPT Policies}\label{sec:WPTOpt}
As aforementioned in Section I, we are interested {in long-term} {\color{black}power} transfer optimization. 
In this section, we first formulate the problem of finding the optimal WPT policy. Assuming the channel statistics are  available, we then derive the optimal solution for the formulated problem, in a closed-form expression. The optimal WPT policy provides a useful insight for finding an effective policy for the general case when the CSI distribution is not available. Finally, based on this insight, we propose a transmission policy that does not require any explicit knowledge of the CSI distribution and determines the beamforming vector in each timeslot based on the observed instantaneous CSI realizations of the current  timeslot and the  transmission history.

\subsection{Problem Formulation}
{As the transmission power from the E-AP} in timeslot $l$ equals $Tr(\bm{x}[l]\bm{x}^H[l])$, the  expected value of the time-averaged transmission power of the E-AP can be written as follows:
\begin{equation}
\bar{Q}_{AP} = \lim_{L \rightarrow \infty}\frac{1}{L}\sum_{l=0}^{L-1}\mathbb{E}[Tr(\bm{x}[l]\bm{x}^H[l])],
\end{equation}
where the expectation is with respect to the randomness of the CSI of the channels. 
%
%
Similarly, the expected value of the time-averaged 
received power at $ \text{E-R}_i $ will be 
\begin{equation}\label{eq: bar_Q_i} 
\bar{Q}_i = \lim_{L \rightarrow \infty}\frac{1}{L}\sum_{l=0}^{L-1}\eta \mathbb{E}[Tr(\bm{W_i}[l]\bm{x}[l]\bm{x}^H[l])],\forall i = 1,...,K.
\end{equation}

An optimal WPT policy of the E-AP aims at maximizing the  power transfer efficiency by maximizing the total received power of the {E-Rs while} satisfying the average and peak power level constraints of the E-AP. 
 Consequently, the problem of finding the optimal WPT policy can be formulated as the following optimization problem:
\begin{maxi!}|l|
 {\{\bm{x}(\bm{H}^{(l)})\}}{\lim_{L \rightarrow \infty}\frac{1}{L}\sum_{l=0}^{L-1}\sum_{i=1}^K\eta\mathbb{E}[Tr(\bm{W}_i[l]\bm{x}[l]\bm{x}^H[l])],}{\label{PEnergyTrans}}{}\label{PEnergyTransObj}
  \addConstraint{ \lim_{L \rightarrow \infty}\frac{1}{L}\sum_{l=0}^{L-1}\mathbb{E}[Tr(\bm{x}[l]\bm{x}^H[l])]\le P_{avg}} \label{equ:PEnergyTransMAXRECV}
\addConstraint{{Tr(\bm{x}[l]\bm{x}^H[l])}\le P_{peak}, \ \forall l \ge 0 }, \label{equ:PEnergyTransP2peak}
\end{maxi!}
%
{where constraints \eqref{equ:PEnergyTransMAXRECV} and \eqref{equ:PEnergyTransP2peak} denote the physical layer limitations on the average and instantaneous transmission power levels of the E-AP, respectively.}

Note that 
{problem formulation \eqref{PEnergyTrans}} is a stochastic optimization problem which is  highly non-trivial and involves some challenges that need to be tackled {appropriately:} First of all, the problem is clearly non-convex due to its  objective function. Moreover, the expectation terms involved in the objective function {and} constraint \eqref{equ:PEnergyTransMAXRECV} do not have any closed-form expression {since} the distributions of the CSI of the channels are not available in practice. 

\subsection{Optimal Power Transfer Policy} \label{subsec:OptimalPolicy}
The following theorem {describes} the optimal solution for {\color{black}problem formulation \eqref{PEnergyTrans}.} The proof is presented in Appendix \ref{app:TheoremOpt}.
\begin{theorem}\label{theo:OptimalWPT}
The following transmission policy maximizes \eqref{PEnergyTransObj} and satisfies constraints \eqref{equ:PEnergyTransMAXRECV} and \eqref{equ:PEnergyTransP2peak}:
In each timeslot, the E-AP  estimates {the CSI of its outgoing links}  and determines the beamforming vector as:
\begin{align}
\bm{x}^*[l] =  \left\{
\begin{array}{cc}
P_{peak}\bm{u}_{max}^{\bm{W'}}[l],\ \   &\lambda_{max}^{\bm{W'}}[l]\ge \lambda_{Th}^{\bm{W'}}, \\
0, & otherwise,
\end{array} \right.
\label{equ:MDPPMaxRecv}
\end{align}
where $\bm{u}_{max}^{\bm{W'}}$ is the eigenvector of matrix $\bm{W'}[l] \triangleq \sum_{i=1}^K \bm{W}_i[l]$ associated with the largest eigenvalue ($\lambda_{max}^{\bm{W'}}[l]$) and
\begin{equation}
\lambda_{Th}^{\bm{W'}}=F^{-1}_{\lambda_{max}^{\bm{W'}}}(1-\frac{P_{avg}}{P_{peak}}), \label{equ:lambaThTheorem3}
\end{equation} 
where $F^{-1}_{\lambda_{max}^{\bm{W'}}}$ is the inverse cumulative distribution function of $\lambda_{max}^{\bm{W'}}$.  
\end{theorem}

The optimal transmission policy, introduced by Theorem \ref{theo:OptimalWPT}, is a two-level policy, in which the E-AP transmits with maximum power when {the quality of the channel is high}; otherwise, it stops transmission. Moreover, when transmitting, the E-AP concentrates the transmission beam toward a virtual E-R with a channel matrix equal to the sum of all the channel matrices. Under this policy, the {power transmission beam} is always biased toward the E-Rs {which have higher quality.} Moreover, {in order to} calculate the optimal threshold in  \eqref{equ:lambaThTheorem3}, the E-AP  needs to know the  distribution of the largest eigenvalue of the  sum of the channel matrices, which may not be available in practice. Although the above issue makes {finding the} optimal policy impractical in many applications, finding the optimal solution can serve as an {upper bound} for the performance of any other policy and {sheds light upon}  the structure of a proper sub-optimal transmission strategy.

\subsection{Near-Optimal Power Transfer Policy}
In this part, based on the Min-Drift-Plus-Penalty (MDPP)  algorithm \cite{Neely2010},
 we propose a near-optimal power transmission policy,  that does not require the CSI distribution.  
The MDPP algorithm is a general {framework for solving stochastic} optimization problems with average constraints. This framework includes a deterministic inner optimization problem that should be addressed for each specific problem formulation, properly. Here, it can be shown that the problem formulation described in equation \eqref{PEnergyTrans} conforms with the MDPP {framework, and hence, in order to propose a near-optimal solution,} it suffices to solve the associated deterministic inner problem. 

The pseudo-code of the proposed solution is presented in Algorithm \ref{alg:RecvMax}. 
The proposed WPT policy only needs the instantaneous CSI realizations and adapts to variations in the CSI distribution.
The  proposed policy follows a similar  two-level transmission strategy as in the optimal solution derived in Section \ref{subsec:OptimalPolicy}. {In this algorithm, variable} $ l $ indicates the timeslot index, {and the process $Z$} represents a virtual queue that captures the deviation of the transmitted power from $P_{avg}$. In fact, the variable $Z[l]$ is an indicator of the {accumulative deviation of the transmission power so far (i.e., up to timeslot l) from 
the allowed transmission power in each timeslot imposed by constraint \eqref{equ:PEnergyTransMAXRECV}.    
Furthermore,} the beamforming vector is  determined in {lines} \ref{alg:s1}-\ref{alg:s2}  of  Algorithm  \ref{alg:RecvMax}.  The parameter $V$ involved in these lines is a control parameter of the MDPP algorithm {\color{black}(for more details on this parameter see \cite{Neely2010}).} We will show via numerical simulations that this parameter maintains a trade-off between {\color{black}the optimality and the convergence time of the algorithm (that is} defined as the number of timeslots that needs to be passed until the  average power constraint is nearly satisfied with a certain bounded deviation gap). 
{Finally,} similar to the optimal solution in Theorem \ref{theo:OptimalWPT}, the beamforming vector in Algorithm \ref{alg:RecvMax} is determined by $\bm{u}_{max}^{\bm{W}'}[l]$, which is the eigenvector  associated with the  largest  {\color{black}eigenvalue of the sum channel 
matrix $\bm{W}'[l]$.} 

{Under the proposed Algorithm \ref{alg:RecvMax}, at the beginning of each timeslot $l$,} the E-AP estimates {\color{black}the CSI of its outgoing links} and calculates the sum channel matrix $\bm{W'}[l]$.  Then, if the largest eigenvalue of $\bm{W'}[l]$, denoted by $\lambda^{\bm{W'}}_{max}[l]$, is greater than $\frac{Z[l]}{V}$, the E-AP will transmit with its maximum power in the direction of the {$\bm{u}_{max}^{\bm{W'}}[l]$; otherwise,} the E-AP will not transmit any power. {Note that this condition (i.e., $\lambda^{\bm{W'}}_{max}[l]\ge \frac{Z[l]}{V}$)} does not require the CSI distribution and  replaces the condition $\lambda^{\bm{W'}}_{max}[l]\ge \lambda^{\bm{W'}}_{Th}$ in the optimal {solution \eqref{equ:MDPPMaxRecv}.} {\color{black}Moreover, it clearly} shows the effects of CSI, virtual queue backlog, and control parameter $V$ on the {transmission policy} in the current timeslot. {For example, if} the quality of the channels is {high in the current timeslot, then with a high probability,  the E-AP will transmit} power to the E-Rs. {\color{black}In addition,} a {larger} value for the virtual queue backlog $Z[l]$ indicates that the {transmission power} deviates much from the average power constraint. {Accordingly, a more power conservative transmission policy should be adopted which} transmits less often. {Furthermore, as $V$ increases, the transmission policy} becomes less sensitive to $Z[l]$. Therefore, the values of $Z[l]$ can {\color{black}increase} without affecting the transmission in the current timeslot, and hence, the convergence time will {\color{black}increase, as well.} Finally, at the end of each timeslot, the virtual queue backlog $ Z $, which {\color{black}is an indicator of} the transmission history, is updated. 

{Note that using the Householder transformations \cite{flannery1992numerical} for the eigenvalue decomposition of $\bm{W}'$, in each timeslot, the computational complexity of both the optimal and near-optimal solutions would be $O(N^3)$, which is polynomial in terms of the number of the E-AP's antennas. \color{black}Finally,} the following theorem {shows} that under the proposed policy, the expectation of the total time-averaged received power, denoted by $\bar{Q}_{PL}^{MDPP}$, {is always} within a bounded distance of the one under the optimal policy, denoted by $\bar{Q}_{PL}^{Opt}$. The proof of this theorem is presented in Appendix \ref{app:NearOptMaxSum}.

\begin{theorem}\label{theo:RecvMaxNearOpt}
The E-AP transmission policy given in Algorithm \ref{alg:RecvMax}:
\begin{enumerate}[label=(\alph*)]
\item
is a feasible solution to problem formulation \eqref{PEnergyTrans} (i.e., it satisfies constraints  \eqref{equ:PEnergyTransMAXRECV} and \eqref{equ:PEnergyTransP2peak}).
\item
yields a total average received power within a maximum distance of $\frac{B}{V}$ from the optimal solution, i.e.,  $\bar{Q}_{PL}^{Opt} \le \bar{Q}_{PL}^{MDPP}\le \bar{Q}_{PL}^{opt}+\frac{B}{V}$, where $B = \frac{1}{2}P_{peak}^2$. 
\end{enumerate}
\end{theorem}
\begin{figure}
\removelatexerror
\begin{algorithm}[H]
\caption{The proposed near-optimal WPT {algorithm.}}\label{alg:RecvMax}
\begin{algorithmic}[1]
\State \textbf{Initialization:} $l\gets 0, Z[0] \gets 0$.
\While{(true)}
\State Estimate $\bm{H_i},\;\forall i = 1,2,\ldots,K$.
\State $\bm{W}_i[l] \gets \bm{H}^H_i[l]\bm{H}_i[l],\;\forall i = 1,2,\ldots,K$.
\State $\bm{W'}[l] \gets \sum_{i=1}^K \bm{W}_i[l]$. \label{alg:s1}
\If{$\lambda_{max}^{{\bm{W}'}}[l]\ge \frac{Z[l]}{V}$} \label{step:BeginningOFWPT}
\State $\bm{x}[l]\gets P_{peak}\bm{u}_{max}^{\bm{W'}}[l]$.
\Else
\State $\bm{x}[l] \gets 0$.
\EndIf \label{alg:s2} \label{step:EndOFWPT}
\State $Z[l+1] \gets \max\{Z[l]+Tr(\bm{x}[l]\bm{x}^H[l])-P_{avg},0\}$. \label{step:VirtualQueue}
\State $l \gets l+1$.
\EndWhile
\end{algorithmic}
\end{algorithm}
\end{figure}
\color{black}
\section{Considering Fairness Among the E-Rs} \label{sec:WPTOptFairness}
{Although the proposed transmission policy in Algorithm \ref{alg:RecvMax} is near-optimal in terms of the} {total received power of the E-Rs, it is} highly biased in favor of those E-Rs that are nearer to the E-AP. This is because {nearer E-Rs will receive} more power  than farther E-Rs if the same amount of power is transmitted toward them. To address this {issue in the design of WPT policy,} in  this section, we aim to ensure fairness among the E-Rs and support their required QoS. For this purpose, we consider a generic network utility function that is concave\footnote{Note that the concavity assumption reduces the {\color{black}difference} between the received {power} of the E-Rs at the cost of reducing the total received power.}, continuous, and non-decreasing  with respect to the average received power of the E-Rs. It is noted that the considered network utility function includes many well-known fairness models, such as max-min fairness, proportional fairness, and $\alpha$-fairness \cite{Omidvar2018Optimal}. Furthermore, we guarantee a  minimum required power for each {\color{black}E-R, denoted by $P_{min}$.} Therefore, the considered  QoS-aware fair WPT problem can be formulated as
\begin{maxi!}|l|
 {\{\bm{x}(\bm{H}^{(l)})\}}{\bar{Q}_\phi \triangleq \phi(\bar{\bm{Q}})}{\label{equ:PFairnessGen}}{}
  \addConstraint{ \lim_{L \rightarrow \infty}\frac{1}{L}\sum_{l=0}^{L-1}\eta\mathbb{E}[Tr(\bm{W}_i[l]\bm{x}[l]\bm{x}^H[l])]\ge P_{min}}  \nonumber
  \addConstraint{\forall i = 1,...,K}\label{equ:PminQGFair}
\addConstraint{ \lim_{L \rightarrow \infty}\frac{1}{L}\sum_{l=0}^{L-1}\mathbb{E}[Tr(\bm{x}[l]\bm{x}^H[l])]\le P_{avg}} \label{equ:PavgQGFair}
\addConstraint{{Tr(\bm{x}[l]\bm{x}^H[l])}\le P_{peak}, \forall l \ge 0 }, \label{equ:PpeakQGFair}
 \end{maxi!}
where $\phi(.)$ is the generic network utility function described above and  $\bar{\bm{Q}}\triangleq (\bar{Q}_1,...,\bar{Q}_K)$ is the aggregated vector of the E-Rs' average received {power}, as defined in \eqref{eq: bar_Q_i}. Constraint \eqref{equ:PminQGFair} guarantees the minimum required power of each E-R. Moreover, 
{same as before, constraints \eqref{equ:PavgQGFair} and \eqref{equ:PpeakQGFair} denote the physical layer limitations on the average and instantaneous transmission power levels of the E-AP, respectively.}

{Note that the formulated problem in \eqref{equ:PFairnessGen} is non-convex and highly non-trivial since the objective function and constraints \eqref{equ:PminQGFair} and \eqref{equ:PavgQGFair} include expectation terms which and do not have any closed-form expressions. To address {\color{black}these challenges} and solve the problem, we use the MDPP technique and propose a policy, named as {\color{black}quality-of-service-aware fair WPT (QF-WPT)}, that maximizes the generic utility function while satisfying the constraints.} 
{\color{black} Note that similar} to the policies in Section \ref{sec:WPTOpt}, the proposed QF-WPT policy follows a  two-level structure and {\color{black}concentrates} the transmission beam  toward a virtual E-R.

The pseudo-code of the proposed policy is presented in Algorithm \ref{alg:RecvGPFFairness}. The E-AP estimates {\color{black}the CSI of its outgoing links} at the beginning of each timeslot and calculates $\bm{W}'[l]$, which is determined as a weighted sum  of the channel matrices of all the E-Rs. The  weights are determined by the virtual queues  ${G}_i,\ Z_i,\  \forall i=1,2,\ldots,K$. The backlogs of $G_i$ and $Z_i$ are more for the E-Rs which have received less power compared to the others, and the E-Rs  which have received less than the minimum required power, respectively. As a consequence, such E-Rs have a {\color{black}higher} weight in the weighted-sum channel matrix and  receive more power in the current timeslot. Then, in lines \ref{step:optimization} and \ref{step:HistoryFairness}, the backlogs of $G_i$'s, which are responsible for ensuring  fairness among the ERs, are updated. It can be easily inferred from line \ref{step:optimization} that the value of $\gamma_i$ has {\color{black}an inverse} relationship with the value of $G_i$. Hence, the backlog of $G_i$ increases more for E-Rs that have less $G_i$ and E-Rs that have received less power in the current timeslot. Such E-Rs will receive more power in the subsequent timeslots as their weights in $\bm{W}'$  increase more compared to the others. Finally, the virtual queues corresponding to the minimum required power and the average transmitted power constraints are updated in lines \ref{step:HistoryMinPower} and \ref{step:HistoryAvgPower}. 

Note that clearly, the optimization problem in line \ref{step:optimization} of the algorithm is convex. Therefore, it can be easily solved using the barrier {\color{black}methods, with} a computational complexity of $O(N\log{(N)})$ \cite{boyd2004convex}. Moreover, using the Householder transformations, the computational complexity of computing
the eigenvalue decomposition of matrix $\bm{W}'$ in line \ref{alg:EigDecomp} of Algorithm \ref{alg:RecvGPFFairness} would be $O(N^3)$ \cite{flannery1992numerical}. Consequently, the total {\color{black}per-iteration} timeslot complexity of our proposed algorithm will be $O(N^3)$. Finally, the following theorem {\color{black}derives the optimality gap} of the proposed QF-WPT policy. The proof of this theorem is presented in Appendix \ref{app:QGFPolicy}.

\begin{theorem}\label{theo:QGFPolicy}
 The QF-WPT policy for the E-AP transmission, described by Algorithm \ref{alg:RecvGPFFairness}:
\begin{enumerate}[label=(\alph*)]
\item
satisfies constraints \eqref{equ:PminQGFair}--\eqref{equ:PpeakQGFair}.
\item
yields {a near-optimal solution  that is within a maximum distance of $\frac{B}{V}$ from the optimal solution, i.e., $\bar{Q}_{\phi}^{Opt}-\frac{B}{V}\le \bar{Q}_{\phi}^{MDPP} \le \bar{Q}_{\phi}^{Opt}$, where $\bar{Q}_{\phi}^{MDPP}$ and $\bar{Q}_{\phi}^{Opt}$ denote  the maximum $\bar{Q}_\phi$ under the QF-WPT policy and the optimal policy, respectively and $B = \frac{2K+1}{2}P_{peak}^2$. }
\end{enumerate}
\end{theorem}


\begin{figure}
\removelatexerror
\begin{algorithm}[H]
\caption{The proposed QoS-aware Fair WPT (QF-WPT) algorithm.}\label{alg:RecvGPFFairness}
\begin{algorithmic}[1]
\State \textbf{Initialization:} $l \gets 0$, $Z_{AP}[0] \gets 0$, $Z_i[0],G_i[0] \gets 0,\ \forall i = 1,2,...,K$.
\While{(true)}
\State Estimate $\bm{H_i}[l],\;\forall i = 1,2,\ldots,K$.
\State $\bm{W}_i[l] \gets \bm{H}^H_i[l]\bm{H}_i[l],\;\forall i = 1,2,\ldots,K$.
\State $\bm{W'}[l] \gets \sum_{i=1}^K (Z_i[l]+{G_i}[l])\bm{W}_i[l]-Z_{AP}[l]\bm{I}$.
\If{$\lambda_{max}^{{\bm{W}'}}[l]\ge 0$}   \label{alg:EigDecomp}
\State $\bm{x}[l] \gets P_{peak}\bm{u}_{max}^{{\bm{W}'}}[l]$.
\Else
\State $\bm{x}[l] \gets 0$.
\EndIf		\label{alg:EigDecompEnd}
\State \multiline{Solve $\min_{\bm{\gamma}} -V\phi(\bm{\gamma})+\sum_{i=1}^K G_i[l]\gamma_i[l],$ where   $\bm{\gamma} \triangleq (\gamma_1[l],...,\gamma_K[l])$}  \label{step:optimization}
\Statex $\qquad \qquad  s.t.  \quad  \gamma_i[l] \le P_{peak},\ \forall i \in \{ 1,\ldots,K\}$. \nonumber
\State \multiline{$G_i[l+1]\gets \max \{G_i[l]+\gamma_i[l]-Tr(\bm{W}_i[l]\bm{x}[l]\bm{x}^H[l]),0\},\;\forall  i = 1,2,...,K$.} \label{step:HistoryFairness}
\State \multiline{$Z_i[l+1] \gets \max\{Z_i[l]+P_{min}-Tr(\bm{W}_i[l]\bm{x}[l]\bm{x}^H[l]),0\},\;\forall i = 1,2,...,K$.} \label{step:HistoryMinPower}
\State $Z_{AP}[l+1] \gets \max\{Z_{AP}[l]+Tr(\bm{x}[l]\bm{x}^H[l])-P_{avg},0\}$.\label{step:HistoryAvgPower}
\State $l\gets l+1$.
\EndWhile
\end{algorithmic}
\end{algorithm}
\end{figure}

\section{Energy Utilization for Wireless Information Transfer}\label{sec:WPCN}
In this section, we consider a scenario in which the E-Rs {utilize} their harvested energy during the WPT phase to successively transmit information to the E-AP during the  energy utilization phase. 
This scenario {\color{black}widely appears} in {\color{black}wireless sensor networks (WSNs)} and IoT networks in which several low-complexity E-Rs {\color{black}rely on} receiving energy from an E-AP to transmit their information back to it {\color{black}(e.g., see \cite{Kang2015} and \cite{Rezaei2019Secrecy}).} 
{\color{black}We aim to find an efficient} wireless power and information transmission policy that maximizes a generic network utility function while guaranteeing a minimum average throughput for each E-R as well as the physical layer constraints for the transmission power of the E-AP.

{\color{black}To formulate the aforementioned problem, first note that} following equation \eqref{equ:InfoOneSlot}, the expected value of the time-averaged throughput for E-R$_i$ can be written as    
\begin{equation}
\bar{D}_i = \lim_{L\to \infty}\frac{1}{L}\sum_{l=0}^{L-1}\mathbb{E}[\tau_u^i[l] \log |\bm{I} + \bm{H}'_i[l] \bm{S}_i[l]\bm{H}_i^{'H}[l]|].
\end{equation}
{\color{black}Therefore, the considered problem can be formulated as follows:}
%
\begin{maxi!}|l|
 {\bm{y}(\bm{H}^{(l)})}{\bar{D}_\phi \triangleq \phi(\bar{\bm{D}})}{\label{PWPCN}}{}
  \addConstraint{\bar{D}_i\ge D_{min},\ \forall i = 1,...,K} \label{equ:QosConstraintWPCN}
  \addConstraint{ \lim_{L \rightarrow \infty}\frac{1}{L}\sum_{l=0}^{L-1}\mathbb{E}[\tau_0[l]Tr(\bm{S}_{AP}[l])]\le P_{avg}} \label{equ:PTRANAVGWPCN}
\addConstraint{{Tr(\bm{S}_{AP}[l])}\le P_{peak}, \ \forall l} \label{equ:PWPCNpeak}
\addConstraint{{\tau_u^i[l] Tr(\bm{S}_i[l])}\le \tau_0[l] Tr(\bm{W}_i[l]\bm{S}_{AP}[l]), \   \forall l, \forall i} \label{equ:EnergyConsumedWPCN}
\addConstraint{\tau_0[l]+\sum_{i=1}^K \tau_u^i[l] = 1,\ \forall l.} \label{equ:WITTauIWPCN}
\end{maxi!}
where $\bm{y}(\bm{H}^{(l)})\triangleq \Big(\bm{S}_{AP}(\bm{H}^{(l)}),\{ \bm{S}_{i}(\bm{H}^{(l)}) \}^{i=1:K}, \bm{\tau} (\bm{H}^{(l)})\Big)$ is the set of optimization variables that are functions of the CSI history of the network until timeslot $l$ (i.e., $\bm{H}^{(l)}$). $\phi(.)$ is a generic concave, continuous, and entrywise non-decreasing fair utility function of the
throughput of the E-Rs, $\bm{S}_{AP}\triangleq \bm{x}\bm{x}^*$, and $\bar{\bm{D}}\triangleq (\bar{D}_1,...,\bar{D}_K)$. Constraint \eqref{equ:QosConstraintWPCN} guarantees {\color{black}the required} QoS (in terms of the minimum average throughput)  for each E-R. 
Moreover, constraints \eqref{equ:PTRANAVGWPCN} and \eqref{equ:PWPCNpeak} are the average and peak power transmission constraints for the E-AP. Furthermore, constraint \eqref{equ:EnergyConsumedWPCN}  ensures that in each timeslot, the consumed energy of each E-R does not exceed its harvested energy.
Finally, constraint \eqref{equ:WITTauIWPCN} {\color{black}guarantees that} the total duration of the WPT phase and the utilization phase {\color{black}in each timeslot equals to one.}

{\color{black}Note that the formulated} problem is non-convex due to its objective function and constraints \eqref{equ:PTRANAVGWPCN} and  \eqref{equ:EnergyConsumedWPCN}. {\color{black}In addition, due to the expectation terms involved,} the objective function and constraints \eqref{equ:QosConstraintWPCN} and \eqref{equ:PTRANAVGWPCN} do not {\color{black}have any} closed-form expressions. {In the rest of this section, we tackle these challenges and propose a near-optimal MDPP-based solution and analyze its performance.} The proposed solution {has} a two-level structure in which the E-AP decides to transmit power or stop transmission based on the CSI quality and the transmission history. In the case of transmission, the E-AP transmits power toward the E-R which has better CSI quality or has transmitted less information  in the previous timeslots. Assuming the E-AP has a large number of antennas compared to the number of the E-Rs, it can generate a sharp beam toward this E-R to transfer all its power to it \cite{Larsson2014MassiveMimo}. Then, this E-R uses the harvested energy to transmit its information with the goal of maximizing its throughput under the fairness and the E-AP's average and peak power level constraints.

Algorithm \ref{alg:ITRecvPropFairness} describes the proposed {\color{black}QoS-aware general fair policy for information transmission (QGF-IT).} In this algorithm, the  transmission history of the E-Rs and the E-AP are captured by virtual queues {\color{black}$G_i$, $Z_i$, $\forall i = 1,...,K$, and $Z_{AP}$.}  At the beginning of each timeslot, the CSI of the E-Rs is estimated. 
Then, the E-R that yields the maximum product of its throughput and summation of queue backlogs (i.e., $f_{obj_i}\triangleq D_i(G_i+Z_i)$)  is found among all the E-Rs. For this purpose, first 
for each E-R, the condition in line \ref{line:CheckCondition} determines whether {\color{black}it is better to transmit power toward this E-R or to save the} transmission power for the subsequent timeslots. 
Noted that this condition will be satisfied for any E-R$_i$ if the queue backlog $Z_{AP}$ is small enough, the summation of the queue backlogs $Z_i$ and $G_i$ is large enough, or the CSI quality of E-R$_i$ is good enough.  This condition is checked for all the E-Rs, and if it is not satisfied for all the E-Rs,
 the E-AP will stop transmission in the current timeslot; otherwise the optimal values of beamforming vectors and sub-timeslots' duration  regarding each E-R are obtained through lines \ref{line:BeginOpt}-\ref{step:fObjFunc}. {\color{black}Accordingly,} the E-R that results in the best objective function (i.e., $f_{obj}$) is chosen in line \ref{line:OptER}, and the optimal values of the beamforming vector (i.e., $\bm{x}[l]$) and the duration of the sub-timeslots (i.e., $\tau_i$'s) corresponding to the chosen E-R are determined in 
lines \ref{line:OptValueFirst}-\ref{line:OptValueSecond}. {\color{black}Finally,} the increments in the queue backlogs $G_i$'s (denoted by $\gamma_i$'s) are obtained by solving the optimization problem in line \ref{step:optimizationGamma}, and
the transmission history is updated in lines \ref{step:AlgGiQueue}-\ref{step:UpdatePx}, accordingly.

Regarding the computational complexity of the proposed algorithm, first note that using Householder transformations \cite{flannery1992numerical}, the {\color{black}singular value decomposition} (SVD)   in line \ref{step:SVDdecomp} of Algorithm \ref{alg:ITRecvPropFairness}  has a total computational complexity of $O(KN^3)$. Moreover, the optimization problem in line \ref{step:optimizationGamma} is convex, and hence, using barrier methods \cite{boyd2004convex}, it can be solved with a computational complexity of $O(N\log{N})$. Accordingly, in each timeslot, the total computational complexity of the proposed algorithm is $O(KN^3)$.
Finally, the following theorem expresses the optimality gap between the proposed solution in Algorithm \ref{alg:ITRecvPropFairness}, denoted by $\bar{D}_{\phi}^{MDPP}$, and the optimal solution of problem formulation \eqref{PWPCN}, denoted by $\bar{D}_{\phi}^{opt}$. The proof of this theorem is presented in Appendix \ref{app:QPF-IT}.
 
\begin{figure}[!t]
\removelatexerror
\begin{algorithm}[H]
\caption{The proposed {\color{black}QoS-aware} general fair algorithm for information transmission (QGF-IT) in WPCNs.} \label{alg:ITRecvPropFairness}
\begin{algorithmic}[1]
\State \textbf{Initialization:} $l \gets 0, Z_{AP}[0] \gets 0$, $Z_i[0], G_i[0] \gets D_{min},\ \forall i = 1,...,K.$
\While{(true)}
\State Estimate $\bm{H}_i[l],\ \forall i = 1,...,K.$
\For{i=1:K} \label{step:LineBeginForLoop}
\State \multiline{Calculate the singular value decomposition (SVD) \label{step:SVDdecomp}
 of ${\bm{H'}_i[l]\sqrt{G_i[l]+Z_i[l]}} = \bm{U}_i[l]\bm{\Theta}_i[l]\bm{V}_i^{H}[l]$. }
\State \multiline{$\beta_i[l] \gets \frac{1}{r_i[l]}(\frac{\lambda_{\bm{W}_i,1}[l]}{G_i[l]+Z_i[l]}-\sum_{j=1}^{r_i[l]}\frac{1}{|\theta_{i_j}[l]|^2})$, where $r_i[l] \triangleq Rank(\bm{H}'_i[l]),\ \theta_{i_j}[l] \triangleq \bm{\Theta_i}[l](j,j)$, and $\lambda_{\bm{W}_i,1}[l]$ is the maximum eigenvalue of $\bm{W}_i[l]$.}  \label{line:Beta}
\State $\alpha_i[l] \gets \frac{1}{r_i[l]}\sum_{j=1}^{r_i[l]}\log{|\theta_{i_j}^2[l]|}+\frac{Z_{AP}P_{peak}}{r_i[l](G_i[l]+Z_i[l])}-1$.  \label{line:Alpha}
\If{($\beta_i[l]e^{\alpha_i[l]}\ge -e^{-1}$)}  \label{line:CheckCondition}
\State $\delta_i[l] \gets \frac{1}{\beta_i[l]}\mathcal{W}(\beta_i[l]e^{\alpha_i[l]})$. \label{line:BeginOpt}
\State \multiline{$\omega_i[l] \gets \frac{\delta_i[l]P_{peak}\lambda_{\bm{W}_i,1}[l]}{(G_i[l]+Z_i[l])\sum_{j=1}^{r_i[l]}\psi_{i_j}[l]}$, where $\psi_{i_j}[l] \triangleq \max{(0,1-\frac{\delta_i[l]}{\theta_{i_j}^2[l]})},\ \forall j=1,...,r_i$.}
\State $\bm{S}_i[l] \gets \frac{G_i[l]+Z_i[l]}{\delta_i[l]}\bm{V}_i[l]\bm{\Psi}_i[l]\bm{V}_i^H[l]$.
\State \multiline{$\tau_{0_i}[l]\gets \frac{1}{1+\omega_i[l]},\tau_u^i[l]\gets \frac{\omega_i[l]}{1+\omega_i[l]}$, $f_{obj_i}[l]\gets(Z_i[l]+G_i[l])D_i[l]$.}  \label{step:fObjFunc}
\Else
\State $\delta_i[l] \gets 0$, $\tau_{0_i}[l] \gets 0$, $\tau_u^i[l] \gets 1$, $f_{obj_i}[l]\gets 0$.
\EndIf
\EndFor  \label{step:LineEndForLoop}
\State $Ind \gets \mathrm{argmax}_{i\in \{ 1,...,K \}}f_{obj_i}[l]$. \label{line:OptER}
\State  $\tau_{0}[l] \gets \tau_{0_{Ind}}[l]$, $\tau_u^{i}[l] \gets 0,\ \forall i \ne Ind$. \label{line:OptValueFirst}
\State \multiline{$\bm{x}[l] \gets \sqrt{P_{peak}}\bm{u}_{\bm{W}_{Ind},1}[l]$, where $\bm{u}_{\bm{W}_{Ind},1}[l]$ is the eigenvector corresponding to the eigenvalue $\lambda_{\bm{W}_{Ind},1}[l]$.} \label{line:OptValueSecond}
\State \multiline{$D_i[l] \gets \tau_u^i[l] \log{|\bm{I}+{\bm{H}'_i[l]}\bm{S}_i[l]{\bm{H}_i^{'H}[l]}|},\ \forall i =  1,...,K$.}  \label{step:ObtainOptimalThr}
\State \multiline{Solve $\min_{\bm{\gamma}} -V\phi(\bm{\gamma})+\sum_{i=1}^K G_i[l]\gamma_i[l],$ where $\bm{\gamma}\triangleq(\gamma_1[l],...,\gamma_K[l]),$}  \label{step:optimizationGamma}
\Statex $\qquad \qquad  s.t.  \quad  \gamma_i[l] \le M\log{(1+\frac{P_{peak}}{\sigma^2})},\ \forall i = 1,...,K$. \nonumber
\State \multiline{$G_i[l+1]\gets \max \{G_i[l]+\gamma_i[l]-D_i[l],0\},\ \forall i = 1,...,K$.} \label{step:AlgGiQueue}
\State \multiline{$Z_i[l+1] \gets \max\{Z_i[l]+D_{min}-D_i[l],0\},\ \forall i = 1,...,K$.} \label{step:AlgZiQueue}
\State \multiline{$Z_{AP}[l+1] \gets \max\{Z_{AP}[l]+\tau_0[l]Tr(\bm{x}[l]\bm{x}^H[l])-P_{avg},0\}$.} \label{step:UpdatePx}
\State $l\gets l+1$.
\EndWhile
\end{algorithmic}
\end{algorithm}
\end{figure}


\begin{theorem}\label{theo:QPF-IT}
The proposed QGF-IT policy described by Algorithm \ref{alg:ITRecvPropFairness}:
\begin{enumerate}
\item
satisfies constraints \eqref{equ:PTRANAVGWPCN}--\eqref{equ:WITTauIWPCN}.
\item
yields a near-optimal solution within a maximum distance of $\frac{B}{V}$ from the optimal solution, i.e., $\bar{D}_{\phi}^{opt}-\frac{B}{V} \le \bar{D}_{\phi}^{MDPP} \le \bar{D}_{\phi}^{opt}$, where $B = \frac{2K+1}{2}P_{peak}^2$.
\end{enumerate}
\end{theorem}

\section{Numerical Results}
\begin{figure}
\centering
  \includegraphics[width=0.9\linewidth]{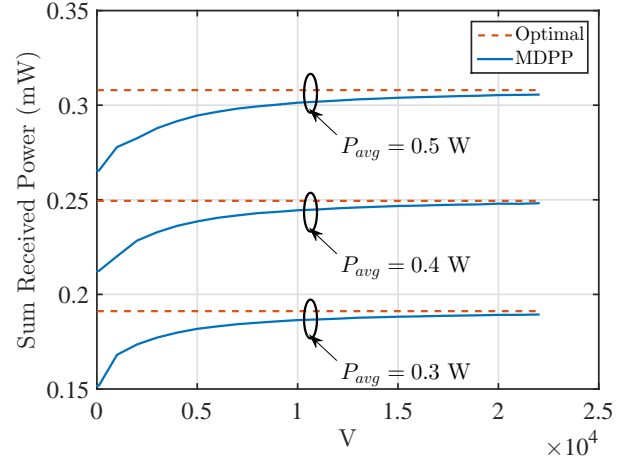}
  \caption{ The average total received power of the
E-R versus the control parameter $V$ for the proposed near-optimal Algorithm \ref{alg:RecvMax} and the optimal solution.}
  \label{fig:NearOptVsParV}
  \end{figure}

\begin{figure}
  \centering
  \includegraphics[width=0.9\linewidth]{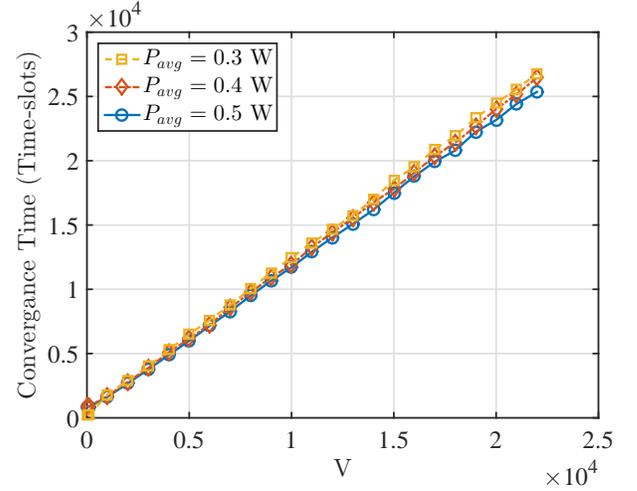}
  \caption{The convergence time of Algorithm \ref{alg:RecvMax} versus the control parameter $V$.}
  \label{fig:ConvTimeVsParV}
\end{figure}

We consider a WPCN 
with one E-AP (located at $(0,0)$ in the two-{\color{black}dimensional} Cartesian space) and $K$ E-Rs, as depicted in Fig. \ref{fig:SysMod}. We set the carrier frequency $f_c = 2.4GHz$, the noise variance of the wireless channels $\sigma^2 = -100dBm$, the E-AP's peak power level $P_{peak} = 2 W$, and the energy conversion efficiency of the E-Rs $\eta = 0.5$. Unless otherwise stated, the numbers of antennas {are considered to be 30 and 4 for the E-AP and each E-R, respectively. Moreover, Rayleigh fading channel model along with a path loss exponent of 3 is considered for all the wireless channels in the network.}
{\color{black}We first consider two E-Rs located at $(1.2,1.2)$ and $(2\sqrt{2},0)$  in the two-dimensional Cartesian space.} Fig. \ref{fig:NearOptVsParV} shows the total average received power of the E-Rs in the optimal and near-optimal (i.e., Algorithm \ref{alg:RecvMax}) solutions versus the control parameter $V$ for several E-AP's average power levels. As can be seen in this figure, as the parameter $V$ increases, the gap between the near-optimal solution and the optimal solution decreases and eventually goes to zero. Furthermore, Fig. \ref{fig:ConvTimeVsParV} shows the {\color{black}convergence time of Algorithm \ref{alg:RecvMax} (which is defined as the total number of timeslots until the deviation of the E-AP's average transmit power from the average power constraint in \eqref{equ:PEnergyTransMAXRECV} falls behind $0.001P_{avg}$) versus the control parameter $V$.} It can be realized from these two figures that as the parameter $V$ increases, the near-optimal solution gets closer to the optimal one with the cost of increasing the convergence time.
%


\begin{figure}
\centering
  \includegraphics[width=0.95\linewidth]{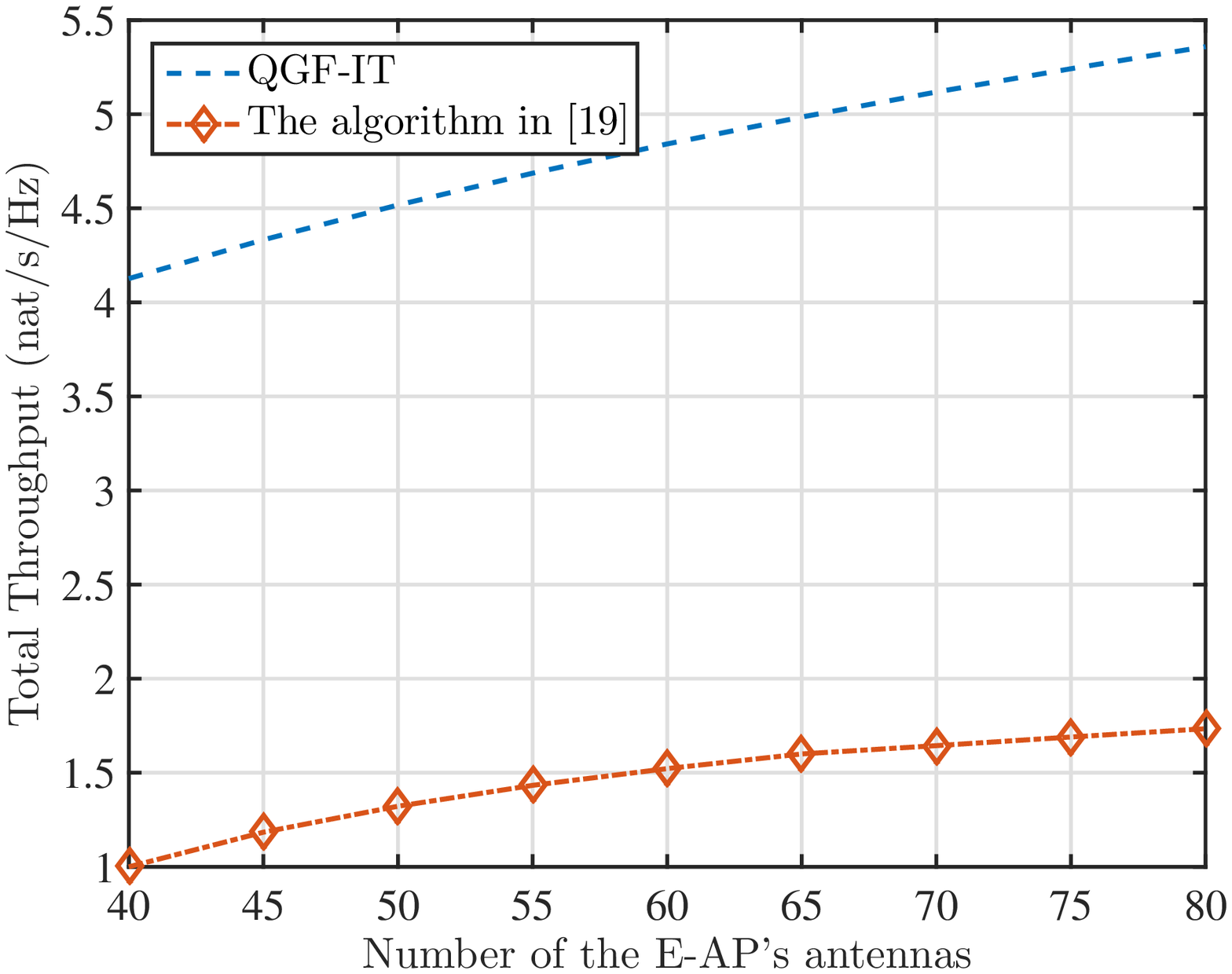}
  \caption{The total throughput of the E-Rs versus the number of the E-AP's antennas.}
  \label{fig:QPFITFirst}
\end{figure}

%
{Next, {\color{black}we evaluate} the performance of the proposed QoS-aware fair policy {\color{black}in Algorithm \ref{alg:RecvGPFFairness} for} two well-known fair utility functions, max-min fairness (MMF) and proportional fairness (PF), {\color{black}and compare it to the performance of the proposed algorithm without considering fairness (Algorithm \ref{alg:RecvMax}). For these simulations, we have considered $P_{avg} = 0.4$ W.}
{\color{black}First, Fig. \ref{fig:FairUsers}} shows the average received power of each E-R versus the distance ratio of the E-Rs, which is defined as $d_r \triangleq \frac{d_f}{d_c}$, where $d_f$ and $d_c$ are the distances of the E-AP to the farther and the closer E-Rs, respectively {(Note that in order to increase $d_r$, we move the farther E-R away
from the E-AP).} 
{\color{black}As can be seen from this figure,} when $d_r=1$, the  E-Rs receive the same amount of power, as expected.
Moreover, it can be verified from this figure 
that, unlike the no-fairness algorithm (i.e., Algorithm \ref{alg:RecvMax}) that allocates almost {\color{black}all the available} power of the E-AP to the closer E-R,
 {\color{black}the proposed MMF QF-WPT} policy allocates an equal amount of power to both the E-Rs irrespective of the {\color{black}value of $d_r$.} However, such an approach may lead to a drastic degradation in the total received power of the E-Rs when the distances between the E-Rs and the E-AP are highly different. In contrast, {\color{black}the proposed PF QF-WPT} policy decreases the received power of the farther E-R smoothly as a function of $d_r$. Hence, the proposed PF QF-WPT policy {results} in a smooth increasing of the gap between the received power of the E-Rs when $d_r$ increases. {\color{black}Moreover, the E-AP guarantees} the required power level of the farther E-R {\color{black}even if} the value of $d_r$ is much {\color{black}greater} than one.

%
%

Fig. \ref{fig:FairUsersTotal} shows the total average received power of the E-Rs ($P_{R,T}$) {\color{black}versus $d_r$}, and compares the {\color{black}proposed} policies with and without fairness. It can be seen from this figure that when considering fairness (either by the proposed MMF QF-WPT or the proposed PF QF-WPT schemes), the total received power of the E-Rs reduces with the increase in the distance ratio. More specifically, under the proposed MMF QF-WPT policy, the total received power of the E-Rs is minimized. Moreover under both the proposed MMF QF-WPT and the proposed PF QF-WPT schemes, the value of  $P_{R,T}$ in our proposed QF-WPT policy, i.e., Algorithm \ref{alg:RecvGPFFairness}, is a monotonically decreasing function of $d_r$.
It can also be {\color{black}verified from this figure} that the {\color{black}proposed PF QF-WPT} policy achieves a good trade-off between the {\color{black}proposed MMF QF-WPT} policy and the no-fairness {\color{black}policy.}

\begin{figure}
\centering
  \includegraphics[width=1\linewidth]{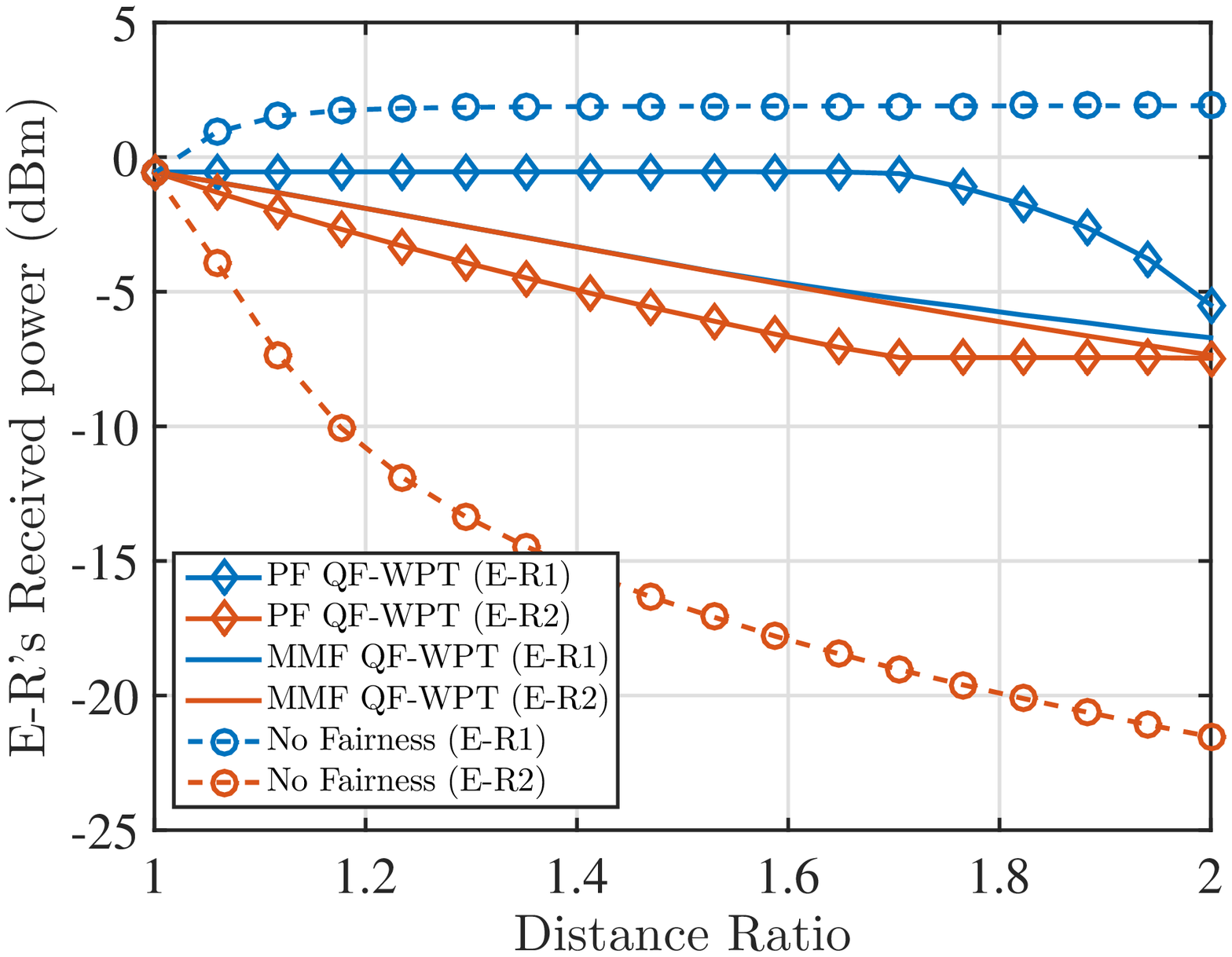}
  \caption{ The average received power of the E-Rs in case of no-fairness, the proposed MMF QF-WPT, and the proposed PF QF-WPT versus the distance ratio of the E-Rs.}
  \label{fig:FairUsers}
\end{figure}
\begin{figure}
  \centering
  \includegraphics[width=1\linewidth]{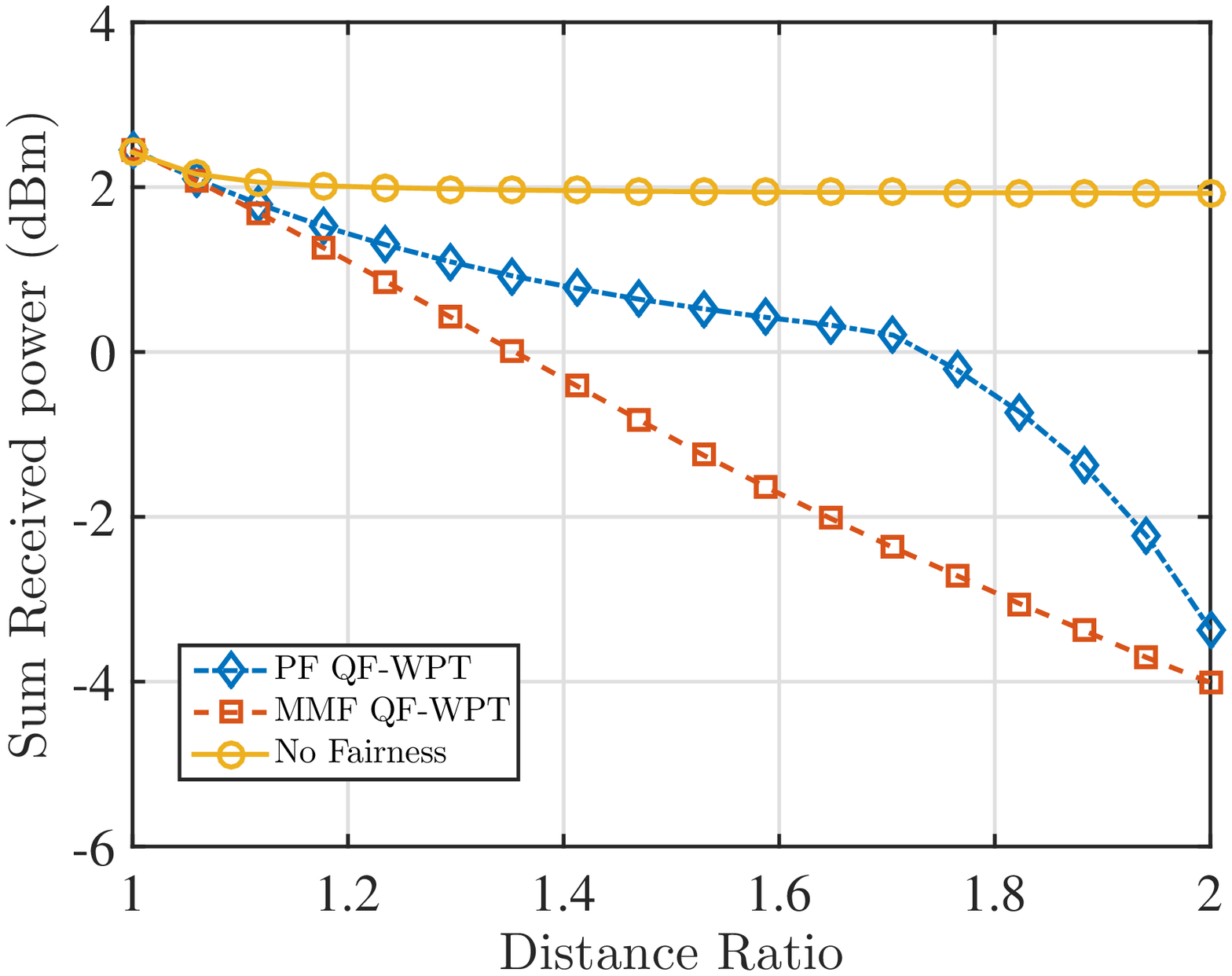}
  \caption{Total average received power of the E-Rs in case of no-fairness, the proposed MMF QF-WPT, and the proposed PF QF-WPT versus the distance ratio  of the E-Rs.}
  \label{fig:FairUsersTotal}
\end{figure}

Finally, Fig. \ref{fig:QPFITFirst} compares the performance of our proposed QGF-IT algorithm to the performance of the algorithm proposed in \cite{Choi2015} (both for the objective function $\phi(\bar{\bm{D}}) = \sum_{i=1}^K \bar{D}_i$) in terms of the total throughput, versus the numbers of the E-AP's antennas.} 
Similar to \cite{Choi2015}, we consider a network topology consisting of one E-AP with $P_{avg}= 0.03$ W and 10 E-Rs (each equipped with one antenna) that are uniformly located at the same distance of $3$ meters from the E-AP.
As can be seen in Fig. \ref{fig:QPFITFirst}, {\color{black}for the same number of the EAP's antennas,} the proposed QGF-IT algorithm {\color{black}always outperforms} the algorithm in \cite{Choi2015} with a significant gap. Moreover, as the number of antennas increases, the performance gap between our proposed algorithm and the algorithm in \cite{Choi2015} increases. This is mainly due to the diversity gain achieved in the uplink information reception of our proposed solution. {\color{black}More specifically,} in each timeslot of the proposed algorithm, the E-AP first {\color{black}utilizes} all of its antennas for power transmission, and then, for information reception; but, 
the work in \cite{Choi2015} has allocated a dedicated antenna for information reception and the remaining antennas for power transfer. Hence, for the same number of the E-AP's antennas, the diversity gain of the proposed algorithm is always greater than of the work in \cite{Choi2015}.   


%

\section{Conclusion}
In this paper, we have studied wireless power transfer as a viable solution to prolong the lifetime of WPCNs. First, we have focused on the problem of maximizing the total average received power of the E-Rs subject to the average and the peak power level constraints of the E-AP. We have formulated the problem as a non-convex stochastic optimization problem and proposed optimal and near-optimal WPT policies to solve this problem. Moreover, we have proved that the proposed near-optimal  solution attains a guaranteed gap to the optimal solution.
Next, we have {\color{black}focused on} the fairness issue among the E-Rs, which is a result of the near-far phenomenon. For addressing this issue, we have proposed a QoS-aware general fair policy for the wireless power transmission from the E-AP to the E-Rs. Finally, we have investigated a generic wirelessly powered communication scenario  in which the E-AP wirelessly transfers power to the E-Rs in the downlink, and the E-Rs utilize their harvested energy to successively transmit their information to the E-AP in the uplink. For this scenario, we have proposed a generic fair policy, referred to as QoS-aware general fair policy for information transmission (QGF-IT), for {\color{black}the} fair transmission of information from the E-Rs. Through various numerical simulations, we have evaluated the performance of the proposed algorithms  and compared them to the state-of-the-art baselines.

\appendices
\section{Proof of Theorem \ref{theo:OptimalWPT}}
\label{app:TheoremOpt}
{\color{black}First, we note that according} to \cite[Theorem 4.5]{Neely2010}, there {\color{black}exists} a stationary solution for the optimization problem defined in equation \eqref{PEnergyTrans}, such that in each timeslot $l$, the decision for the transmission signal of the E-AP (i.e., $\bm{x}$) is only a function of $\bm{H}[l]$ and is independent of the transmission history. Hence, considering only the stationary solutions, we can omit the timeslot index $l$ and the time-averages in equations \eqref{PEnergyTransObj} and \eqref{equ:PEnergyTransMAXRECV} and rewrite the optimization problem as follows: 
\begin{maxi!}|l|
 {\{\bm{x}(\bm{H})\}}{\mathbb{E}[P_xTr(\bm{W}'\bm{\tilde{x}}\bm{\tilde{x}}^H)]}{\label{P2A}}{}\label{equ:P2ObjFunc}
\addConstraint{ \mathbb{E}[P_x]\le P_{avg}} \label{equ:P2ConstrPavg}
\addConstraint{P_x\le P_{peak}}
\addConstraint{\lVert \bm{\tilde{x}}\rVert_2 = 1}, \label{equ:Norm2Eq}
 \end{maxi!}
{where $\bm{x}=P_x \bm{\tilde{x}}$. The problem defined in equation \eqref{P2A} can be solved for $\bm{\tilde{x}}$  independent of the value of $P_x$. Specifically, for a given $\bm{W}'$ and all values of $P_x$, the term $P_xTr(\bm{W}'\bm{\tilde{x}}\bm{\tilde{x}}^H)$ is maximized with respect to $\bm{\tilde{x}}$ at $\bm{\tilde{x}} = \bm{u}_{max}^{\bm{W}'}$ \cite{zhang2013mimo}, and the maximum value equals $P_x \lambda_{max}^{\bm{W}'}$. Now, we set $\bm{\tilde{x}} = \bm{u}_{max}^{\bm{W}'}$ and show that
\begin{align}
P_x =  \left\{
\begin{array}{cc}
P_{peak},\ \   &\lambda_{max}^{\bm{W}'}[l]\ge \lambda_{Th}^{\bm{W}'}, \\
0, & otherwise,
\end{array} \right.
\label{equ:MDPPMaxPxRecv}
\end{align}
maximizes \eqref{equ:P2ObjFunc}. Note that $P_x$ defined in equation \eqref{equ:MDPPMaxPxRecv} satisfies constraint \eqref{equ:P2ConstrPavg} as follows: }
\begin{align*}
\mathbb{E}[P_{x}]&=P_{peak}P(\lambda_{max}^{w'}\ge \lambda_{Th}^{w'})=P_{peak}(1-F_{\lambda_{max}^{w'}}(\lambda_{Th}^{w'}))\\ &=P_{avg}.
\end{align*}
{Consider an alternative policy, denoted by $\bm{\bar{x}}$, for the transmission signal that satisfies constraint \eqref{equ:P2ConstrPavg}. We have, }
\begin{align*}
&\mathbb{E}[\lambda_{max}\bar{P}_x]-\mathbb{E}[\lambda_{max}P_x]\\&=P(\lambda_{max}\ge\lambda_{Th})\mathbb{E}[\lambda_{max}\bar{P}_x|\lambda_{max}\ge\lambda_{Th}] \\&+P(\lambda_{max}<\lambda_{Th})\mathbb{E}[\lambda_{max}\bar{P}_x|\lambda_{max}<\lambda_{Th}]\\&-P(\lambda_{max}\ge\lambda_{Th})\mathbb{E}[\lambda_{max}P_x|\lambda_{max}\ge\lambda_{Th}]\\
& = P(\lambda_{max}\ge\lambda_{Th})\mathbb{E}[\lambda_{max}(\bar{P}_x-P_{peak})|\lambda_{max}\ge\lambda_{Th}]\\&+P(\lambda_{max}<\lambda_{Th})\mathbb{E}[\lambda_{max}\bar{P}_x|\lambda_{max}<\lambda_{Th}] \\
&\le \lambda_{Th}\bigg(P(\lambda_{max}\ge\lambda_{Th})(\mathbb{E}[(\bar{P}_x-P_{peak})|\lambda_{max}\ge\lambda_{Th}])\\&+P(\lambda_{max}<\lambda_{Th})\mathbb{E}[\bar{P}_x|\lambda_{max}<\lambda_{Th}]\bigg) \\
 &= \lambda_{Th}(\mathbb{E}[\bar{P}_x]-\mathbb{E}[P_x]) = \lambda_{Th}(\mathbb{E}[\bar{P}_x]-P_{avg}) \le 0,
\end{align*}
{where $\bar{P}_x$ denotes the transmission power under $\bm{\bar{x}}$. It shows that  the value of the objective function under the alternative policy is always less than or equal to the one under the proposed policy. This completes the proof of Theorem \ref{theo:OptimalWPT}. }
\section{Proof of Theorem \ref{theo:RecvMaxNearOpt}}\label{app:NearOptMaxSum}
Algorithm \ref{alg:RecvMax} is based on the MDPP technique, which uses the Lyapunov optimization method. The following definitions are considered for quadratic Lyapunov function and Lyapunov drift, respectively: 
\begin{equation}
\mathbb{L}({Z}[l]) \triangleq \frac{1}{2} Z^2[l],
\end{equation}
\begin{equation}
\Delta({Z}[l]) \triangleq \mathbb{E}[(\mathbb{L}({Z}[l+1])-\mathbb{L}({Z}[l]))|{Z}[l]].
\end{equation}
Let us define the drift-plus-penalty function as 
\begin{equation}\label{equ:DPP}
\Delta({Z}[l]) + V\mathbb{E}\big[-\sum_{i=1}^K Q_{i}[l]|{Z}[l]\big], 
\end{equation}  
 where $Q_{i}[l]=Tr(\bm{W}_i[l]\bm{x}[l]\bm{x}^H[l])$.  The first term in equation \eqref{equ:DPP} is a measure of the expected total backlog increment in the virtual queue,
and {the second term is negative of the expected received power of all the E-R's, where both are conditional expectation given the current queue backlog (i.e., $Z[l]$). The intuition behind the MDPP technique is to propose a proper policy that minimizes this function. As a result, this policy maximizes the total received power and reduces the length of the virtual queue backlog.}
We derive an upper bound for the drift-plus-penalty function as follows:
\begin{align}
&Z^2[l+1]\stackrel{(a)}{\le} Z^2[l]+ \alpha_{AP}^2[l]+2Z[l] \alpha_{AP}[l] \nonumber \\  \Rightarrow &\Delta({Z}[l]) \le Z[l]\mathbb{E}[\alpha_{AP}[l] |{Z}[l]] + \frac{1}{2} \mathbb{E}[\alpha_{AP}^2|{Z}[l]] \nonumber \\
\Rightarrow &\Delta({Z}[l]) + V\mathbb{E}\big[-\sum_{i=1}^KQ_{i}[l]|{Z}[l]\big] \nonumber \\  \le &B+ V\mathbb{E}\big[-\sum_{i=1}^KQ_{i}[l]|{Z}[l]\big] + Z[l]\mathbb{E}[  \alpha_{AP} [l] |{Z}[l]]], \label{equ:LyapunovIneq}
\end{align}
where $\alpha_{AP} [l] \triangleq   Tr(\bm{x}[l]\bm{x}^H[l]) - P_{avg}$, $B \triangleq {1\over 2}  P_{peak}^2$, and $(a)$ results from the virtual queue update equation (line \ref{step:VirtualQueue} of Algorithm \ref{alg:RecvMax}).
{The transmitted signal at the E-AP has a maximum power of $P_{peak}$, so $\frac{1}{2}\mathbb{E}[\alpha_{AP}[l] ^2|\mathbf{Z}[l]] \le \frac{1}{2}P_{peak}^2 = B$. The ratio $\frac{B}{V}$ is the optimality gap of the proposed solution as mentioned in property (b) of Theorem \ref{theo:RecvMaxNearOpt}.}

According to the Lyapunov optimization theorem \cite[Theorem 4.8]{Neely2010}, a feasible policy which minimizes the right hand side (RHS) of equation \eqref{equ:LyapunovIneq} in each timeslot satisfies the properties (a) and (b) of Theorem \ref{theo:RecvMaxNearOpt}. {Hence, in order to prove Theorem \ref{theo:RecvMaxNearOpt}, it suffices to show that Algorithm \ref{alg:RecvMax} minimizes the RHS of equation \eqref{equ:LyapunovIneq}. Specifically, we show that Algorithm \ref{alg:RecvMax} solves the following problem}
\begin{mini!}|l|
 {\bm{x}[l] }{ -V\sum_{i=1}^K {Q_{i}}[l]+Z[l]\alpha_{AP}[l]}{\label{equ:MDPPMinTran}}{}\label{equ:MDPPMinTranObj}
  \addConstraint{  Tr(\bm{x}[l]\bm{x}^H[l])\le P_{peak}}. 
 \end{mini!}

{Using the definitions for $\alpha_{AP}[l]$ and $Q_i[l]$ in equation \eqref{equ:receivedEnergyEq2}, the problem defined in equation \eqref{equ:MDPPMinTran} can be rewritten as}
\begin{maxi!}|l|
 {\bm{x}[l] }{Tr((\bm{W}'[l]-\frac{Z[l]}{V}\bm{I})\bm{x}[l]\bm{x}^H[l])}{\label{equ:MDPPMinTran2}}{}
  \addConstraint{  Tr(\bm{x}[l]\bm{x}^H[l])\le P_{peak}}. 
 \end{maxi!}
where $\bm{W'}[l] \triangleq \sum_{i=1}^K \bm{W}_i[l]$. The solution of {\color{black}optimization problem formulation \eqref{equ:MDPPMinTran2}} is \cite{zhang2013mimo} 
\begin{align}\label{equ:XlNearOptimal}
\bm{x}[l] =  \left\{
\begin{array}{cc}
P_{peak}\bm{u}_{max}^{\bm{W'}}[l],\ \   &\lambda_{max}^{\bm{W'}}[l]\ge \frac{Z[l]}{V}, \\
0, & otherwise.
\end{array} \right.
\end{align}

{It can be verified that the transmission signal of the E-AP (i.e., $\bm{x}$)  calculated in lines \ref{step:BeginningOFWPT} to \ref{step:EndOFWPT} of Algorithm \ref{alg:RecvMax} follows the same rule as in equation \eqref{equ:XlNearOptimal}. This completes the proof of Theorem \ref{theo:RecvMaxNearOpt}.}

\section{Proof of Theorem \ref{theo:QGFPolicy}}\label{app:QGFPolicy}
{We solve the optimization problem defined in equation \eqref{equ:PFairnessGen} 
using the MDPP approach. However, this approach is not directly applicable to this problem since the  objective function of the problem 
is a function of  the time-averaged received power, which does not conform to the standard MDPP framework. 
Accordingly, we follow the same approach as in \cite[Chapter 5]{Neely2010} and introduce the vector of slack variables $\bm{\gamma}[l]=(\gamma_1[l],...,\gamma_K[l])$ to convert the problem from maximizing  a utility function of time averages to maximizing a time average of the utility function.}
It has been shown in \cite[Chapter 5]{Neely2010} that the optimal solution of the modified problem is the same as the original problem. The modified optimization problem is as follows: 
\begin{maxi!}|l|
 {\{ \bm{y}(\bm{H}^{(l)}) \}}{\overline{\phi(\bm{\gamma})}}{\label{P3'}}{} \label{equ:objectiveModifApp}
\addConstraint{\bar{\gamma}_i\le \bar{Q}_i,\ \forall i = 1,...,K} \label{equ:ModifiedProblemConstApp} 
  \addConstraint{ \lim_{L \rightarrow \infty}\frac{1}{L}\sum_{l=0}^{L-1}\eta\mathbb{E}[Tr(\bm{W}_i[l]\bm{x}[l]\bm{x}^H[l])]\ge P_{min}}  \label{equ:PminQGFairApp}
  \addConstraint{ \lim_{L \rightarrow \infty}\frac{1}{L}\sum_{l=0}^{L-1}\mathbb{E}[Tr(\bm{x}[l]\bm{x}^H[l])]\le P_{avg}} \label{equ:ModifiedProblemPavgApp} 
\addConstraint{{Tr(\bm{x}[l]\bm{x}^H[l])}\le P_{peak}, \forall l \ge 0 }, \label{equ:ModifiedProblemPpeakApp} 
 \end{maxi!} 
where $\bm{y}(\bm{H}^{(l)})\triangleq \Big(\bm{x}(\bm{H}^{(l)}),\bm{\gamma}(\bm{H}^{(l)})\Big)$ and 
\begin{align*}
\overline{\phi(\bm{\gamma})}&\triangleq\lim_{L \rightarrow \infty}\frac{1}{L}\sum_{l=0}^{L-1} \mathbb{E}[ \phi(\gamma_1[l],...,\gamma_K[l])   ],\\ 
 \bar{\gamma}_i&\triangleq\lim_{L \rightarrow \infty}\frac{1}{L}\sum_{l=0}^{L-1}\mathbb{E}[\gamma_i[l]]. 
\end{align*}
The modified optimization problem is a time-averaged stochastic optimization problem similar to  {\color{black}optimization problem formulation \eqref{PEnergyTrans}.} Hence, similar to Appendix \ref{app:NearOptMaxSum}, {we have to define the virtual queues $G_i$, $Z_i, \forall i=1,...,K$, and $Z_{AP}$ corresponding to constraints \eqref{equ:ModifiedProblemConstApp}, \eqref{equ:PminQGFairApp}, and \eqref{equ:ModifiedProblemPavgApp}, respectively. Then, the MDPP approach suggests that a policy which solves the following problem in each timeslot satisfies the properties (a) and (b) of Theorem \ref{theo:QGFPolicy}.}
 \begin{mini!}|l|
 {\bm{\gamma},\bm{x}[l] }{ -V\phi(\bm{\gamma})+Z_{AP}[l](Tr(\bm{x}[l]\bm{x}^H[l])-P_{avg})}{\label{equ:MDPPMaxRec}}{} \nonumber
 \breakObjective {+G_i[l](\gamma_i[l] - Q_i[l])} \nonumber
  \breakObjective {+Z_i[l](P_{min}-Q_i[l])} \label{equ:MDPPMaxRecObj}
  \addConstraint{  Tr(\bm{x}[l]\bm{x}^H[l])\le P_{peak}}. 
 \end{mini!}
{Now, we show that Algorithm \ref{alg:RecvGPFFairness} solves problem formulation \eqref{P3'} in each timeslot.} This problem can be decoupled to two optimization subproblems. The optimal value of $\bm{\gamma}$ is obtained via solving the following optimization problem
 \begin{mini!}|l|
 {\bm{\gamma}}{ -V\phi(\bm{\gamma})+\sum_{i=1}^K G_i[l]\gamma_i[l] }{\label{equ:MDPPMaxRecGamma}}{}\label{equ:MDPPMaxRecObj}
  \addConstraint{ 0\le \gamma_i[l]\le P_{peak},\ \forall i \in \{ 1,...,K\}}. 
 \end{mini!}
Noted that in the above formulation, it is considered that the maximum of $\gamma_i[l]$ in each timeslot is $P_{peak}$, which is the maximum possible received power. 
The optimization subproblem to find the optimal value of $\bm{x}[l]$ is
 \begin{mini!}|l|
 {\bm{x}[l] }{ -\sum_{i=1}^K(G_i[l]+Z_i[l])Tr(\bm{W}_i[l]\bm{x}[l]\bm{x}^H[l]) \nonumber }{\label{equ:MDPPMaxRecGamma}}{}
  \breakObjective {+Z_{AP}[l]Tr(\bm{x}[l]\bm{x}^H[l])} \label{equ:MDPPMaxRecObj}
  \addConstraint{  Tr(\bm{x}[l]\bm{x}^H[l])\le P_{peak}}. 
 \end{mini!}
The optimal solution of this problem is \cite{zhang2013mimo} 
\begin{align}\label{equ:xSolutionGenFairWPT}
\bm{x}[l] =  \left\{
\begin{array}{cc}
P_{peak}\bm{u}_{max}^{\bm{W}'}[l],\ \   &\lambda_{max}^{\bm{W'}}[l]\ge 0, \\
0, & otherwise,
\end{array} \right.
\end{align}
where $\bm{W}'[l]\triangleq \sum_{i=1}^K (Z_i[l]+{G_i}[l])\bm{W}_i[l]-Z_{AP}[l]\bm{I}$. {Now, it can be verified that Algorithm \ref{alg:RecvGPFFairness} solves problem formulation \eqref{P3'} in each timeslot. The policy for determining $\bm{\gamma}$, which is in line \ref{step:optimization} of Algorithm \ref{alg:RecvGPFFairness},  follows optimization problem formulation \eqref{equ:MDPPMaxRecGamma} and the policy for determining $\bm{x}[l]$, which is in lines \ref{alg:EigDecomp} to \ref{alg:EigDecompEnd} of Algorithm \ref{alg:RecvGPFFairness}, follows equation \eqref{equ:xSolutionGenFairWPT}. This completes the proof of Theorem \ref{theo:QGFPolicy}.}

\section{Proof of Theorem \ref{theo:QPF-IT}}\label{app:QPF-IT}
{\color{black}Problem formulation \eqref{PWPCN}} is a function of the time-averaged received throughput. With similar arguments to Appendix \ref{app:QGFPolicy}, the slack variable vector $\bm{\gamma}[l]$ = ($\gamma_1[l]$,...,$\gamma_K[l]$) is introduced to convert the optimization problem from optimizing a function of time averages to optimizing time average. 
The modified problem can be written as follows:
\begin{maxi!}|l|
 {\bm{y}(\bm{H}^{(l)})}{ \overline{\phi(\bm{\gamma})}}{\label{P2GeneralFair}}{}
    \addConstraint{\bar{D}_i\ge \bar{\gamma}_{i}},\ \forall i \in \{ 1,...,K\} \label{equ:AuxilaryVarGPF}
  \addConstraint{\bar{D}_i\ge D_{min}},\ \forall i \in \{ 1,...,K\} \label{equ:QosConstraintGPF}
  \addConstraint{ \lim_{L \rightarrow \infty}\frac{1}{L}\sum_{l=0}^{L-1}\mathbb{E}[\tau_0[l]Tr(\bm{S}_{AP}[l])]\le P_{avg}} \label{equ:MAXRECVPTRANAVGGPF}
\addConstraint{{Tr(\bm{S}_{AP}[l])}\le P_{peak}, \ \forall l  } \label{equ:P2peakGPF}
\addConstraint{\tau_u^i[l] Tr(\bm{S}_i[l])\le \tau_0[l] Tr(\bm{W}_i[l]\bm{S}_{AP}[l]), \ \forall l,\forall i 
} \label{equ:EnergyConsumedGPF}
\addConstraint{\tau_0[l]+\sum_{i=1}^K \tau_u^i[l] = 1,  \ \forall l}, \label{equ:WITTauI}
\end{maxi!}
where 
\begin{equation*}
\bm{y}(\bm{H}^{(l)})\triangleq \Big(\bm{\gamma}(\bm{H}^{(l)}), \bm{S}_{AP}(\bm{H}^{(l)}),\{ \bm{S}_{i}(\bm{H}^{(l)}) \}, \bm{\tau} (\bm{H}^{(l)})\Big). 
\end{equation*}
We have to define the virtual queues $Z_{AP}$, $Z_{i}$'s, and $G_i$'s, $\forall i=1,...,K$ corresponding to  constraints \eqref {equ:AuxilaryVarGPF}, \eqref{equ:QosConstraintGPF}, and \eqref{equ:MAXRECVPTRANAVGGPF}, respectively. Then, the following deterministic optimization problem must be solved in each timeslot to obtain the near-optimal solution: 

\begin{mini!}|l|
 {\bm{\gamma},\bm{S}_{AP},\{\bm{S}_i \},\bm{\tau}}{-V\phi(\bm{\gamma})+ Z_{AP}(\tau_0Tr(\bm{S}_{AP})-P_{avg})\nonumber}{\label{P2Stochastic}}{}
 \breakObjective {+\sum_{i=1}^K Z_{i}(D_{min} - D_i)}\nonumber
\breakObjective { + \sum_{i=1}^K G_i(\gamma_i-D_i)}
  \addConstraint{{Tr(\bm{S}_{AP})}\le P_{peak}} 
\addConstraint{{\tau_u^i Tr(\bm{S}_i)}\le \tau_0 Tr(\bm{W}_i\bm{S}_{AP})} \label{equ:vons:S}
\addConstraint{\tau_0+\sum_{i=1}^K \tau_u^i = 1 }. \label{Constr:No}
\end{mini!}
Note that in the above formulation, the slot index ($l$) is omitted for brevity. The above problem is non-convex due to its objective function and constraint \eqref{equ:vons:S}. To resolve this issue, we introduce slack variables $\bm{S}_i^{'}  = \tau_u^i \bm{S}_i, \forall i = 1,...,K$ and $\bm{S}_{AP}^{'}  = \tau_0 \bm{S}_{AP}$ and reformulate the problem as follows: 
\begin{mini!}|l|
 {\bm{\gamma},\bm{S'}_{AP},\{\bm{S'}_i \},\bm{\tau}}{-V\phi(\bm{\gamma})+ Z_{AP}(Tr(\bm{S'}_{AP})-P_{avg}) \nonumber}{\label{Prob:PMainInEachSlot}}{}\label{Prob:PMainInEachSlotObj}
 \breakObjective {+\sum_{i=1}^K Z_{i}(D_{min} - D_i)}\nonumber
\breakObjective { + \sum_{i=1}^K G_i(\gamma_i-D_i)}
  \addConstraint{{Tr(\bm{S'}_{AP})}\le \tau_0P_{peak}} 
\addConstraint{{Tr(\bm{S'}_i)}\le  Tr(\bm{W}_i\bm{S'}_{AP})} \label{equ:vons:SConvex}
\addConstraint{\tau_0+\sum_{i=1}^K \tau_u^i = 1 }. \label{Constr:No}
\end{mini!}
Note that $D_i = \tau_u^i \log{|\bm{I}+{\bm{H}_i^{'}}\frac{\bm{S}'_i}{\tau_u^i}{\bm{H}_i^{'H}[l]}|}$ is a perspective of the function $\log{|\bm{I}+{\bm{H}_i^{'}}\bm{S}'_i{\bm{H}_i^{'H}[l]}|}$, hence is concave \cite{boyd2004convex}. In addition, $\phi(\bm{\gamma})$ is also concave. As an immediate result, the objective function \eqref{Prob:PMainInEachSlotObj} is convex. Since all the constraints are linear, the optimization problem \eqref{Prob:PMainInEachSlot} is convex. Moreover, it is easy to verify that the Slater's qualification condition holds for an achievable $D_{min}$. Therefore, we can solve this problem by solving its dual problem, which can be written as
\begin{maxi!}
{\bm{\delta},\zeta,\xi}{\min_{\bm{\gamma},\{\bm{S}'_i \},\bm{S}'_{AP},\bm{\tau}}\ \mathcal{L}(\bm{\gamma},\{\bm{S}'_i \},\bm{S}'_{AP},\bm{\tau},\bm{\delta},\zeta,\xi)}{\label{equ:DualMaxMinInfo}}{} \label{equ:DualMaxMinInfoObj}
\addConstraint{\delta_i,\xi \ge 0,\ \forall i \in \{ 1,...,K \},}
\end{maxi!}
where $\bm{\delta} \triangleq (\delta_1,...,\delta_K)$ and 
\begin{align}\label{equ:DualOptimiaaio}
\mathcal{L}&\triangleq -V\phi(\bm{\gamma}) +\sum_{i=1}^K \bigg[Z_i(D_{min}-D_i) + G_i(\gamma_i-D_i) \nonumber \\ &+\delta_i \big( Tr(\bm{S}'_i)-Tr(\bm{W}_i\bm{S}'_{AP}) \big)\bigg] \nonumber +Z_{AP}(Tr(\bm{S}'_{AP})-P_{avg})\\ &+ \zeta(\tau_0+\sum_{i=1}^K \tau_u^i  -1)+\xi(Tr(\bm{S}'_{AP})-\tau_0P_{peak}).
\end{align}

The alternating optimization method is used to solve {\color{black}optimization problem formulation \eqref{equ:DualMaxMinInfo}.} First, the variables $\xi$, $\gamma_i$'s, $\bm{S}'_{AP}$ and $\bm{S}'_i$'s are optimized. Then, their optimal solution is put into the optimization problem formulation \eqref{equ:DualMaxMinInfo}, and the optimal values for the remaining parameters are obtained. 

For $\xi$, it is easy to verify that if $0<\tau_0<1$, then we must have $\xi = \frac{\zeta}{ P_{peak}}$. To obtain the optimal values of $\gamma_i$'s, 
the following optimization problem must be solved:
\begin{mini!}|l|
 {\bm{\gamma}}{ -V\phi{(\bm{\gamma})}+\sum_{i=1}^K{G_i}\gamma_i}{\label{P2StochasticGammaPart}}{}
  \addConstraint{\gamma_i \le D_{max},\ \forall i \in \{ 1,...,K \},}
\end{mini!}
where $D_{max}$ is an upperbound of the maximum throughput of all the E-Rs (a rough approximation is $M\log{(1+\frac{P_{peak}}{\sigma^2})}$).
The optimization of this problem depends on $\phi(.)$ and is done in line \ref{step:optimizationGamma} of Algorithm \ref{alg:ITRecvPropFairness}.
Next, we consider the optimization problem regarding $\bm{S}'_{AP}$, which can be written as follows: 
\begin{mini!}|l|
 {\bm{S}{'}_{AP}}{Tr\bigg(\big((Z_{AP}+\xi)\bm{I} - \sum_{i=1}^K\delta_i \bm{W}_i\big)\bm{S}'_{AP}\bigg).}{\label{P2StochasticSBasePart}}{} 
\end{mini!}
The optimal solution of this problem is \cite{zhang2013mimo}
\begin{align}
\bm{S}^{'}_{AP} =  \left\{
\begin{array}{cc}
{\tau_0P_{peak}}\bm{u}_{\bm{B},1}\bm{u}^H_{\bm{B},1},\ \   &\lambda_{\bm{B},1}\ge Z_{AP}+\xi, \\
0, & otherwise,
\end{array} \right.
\label{equ:SBSPrimeFormula}
\end{align}
where $\bm{B}\triangleq \sum_{i=1}^K\delta_i \bm{W}_i$ and  $\lambda_{\bm{B},1}$ and $\bm{u}_{\bm{B},1}$ are its maximum eigenvalue and the corresponding eigenvector. {We assume that the E-AP has a large number of antennas compared to the total number of all the ERs' antennas\footnote{\color{black}As in practical scenarios, each ER has a low number of antennas, this assumption is almost the same as the E-AP has a large number of antennas compared to the number of the ERs. Though, the obtained solution may be suboptimal for the case that $N$ is comparable to $KM$.} (i.e., $N \gg KM$), so the eigenvectors of matrix $\bm{B}$ is the union of the eigenvectors of $\bm{W}_i$'s \cite{Larsson2014MassiveMimo}}. Hence, $\bm{u}_{\bm{B},1}$ is the eigenvector of one of the $\bm{W}_i$'s, and the E-AP transfers power in the direction of the corresponding E-R. We consider $K$ cases regarding the E-R who receives power in the current timeslot and then, choose the one which has the minimum value of the objective function \eqref{equ:DualMaxMinInfoObj}. Without loss of generality, we assume that E-R$_i$ receives the power and formulate the corresponding subproblem for $\bm{S}'_i$ as follows: 
 \begin{mini!}|l|
 {\bm{S}'_{i}}{-(Z_i+G_i)\tau_u^i\log |\bm{I} + \bm{H}'_i[l] \frac{\bm{S}'_i[l]}{\tau_u^i[l]}\bm{H}_i^{'H}[l]|\nonumber}{\label{P2StochasticSNodesPartSp}}{} 
 \breakObjective{+\delta_i Tr(\bm{S}'_i).}
\end{mini!}
The slack variable $\bm{S}''_i \triangleq \frac{\delta_i}{G_i+Z_i}\frac{\bm{S}'_i}{\tau_u^i}$ is introduced, and the above problem is reformulated as follows:
\begin{maxi!}|l|
 {\bm{S}''_{i}}{\log |\bm{I} + \frac{\bm{H}''_i}{\sqrt{\delta_i}} {\bm{S}''_i} \frac{\bm{H}^{''H}_i}{\sqrt{\delta_i}}|- Tr(\bm{S}''_i),}{\label{P2StochasticSNodesPartSp2}}{} 
\end{maxi!}
where $\bm{H}''_i \triangleq {\bm{H}'_i\sqrt{G_i+Z_i}}$. To obtain the optimal $\bm{S}''_i$, we calculate the SVD decomposition of $\bm{H}''_i = \bm{U}_i\bm{\Theta}_i\bm{V}_i^H,$ where $\bm{U}_i \in \mathbb{C}^{N \times N}$ and $\bm{V}_i \in \mathbb{C}^{r_i \times r_i}$ ($ r_i$ is the rank of $\bm{H}''_i$) are singular left and right vectors of $\bm{H}''_i$, respectively. 
Now, the optimal solution of the problem defined in equation \eqref{P2StochasticSNodesPartSp2} can be written as follows \cite{cover2012elements}:
\begin{equation}
\bm{S}^{''}_i = \bm{V}_i\bm{\Psi}_i\bm{V}_i^H \Rightarrow \bm{S}^{'}_i = \tau_u^i \frac{G_i+Z_i}{\delta_i}\bm{V}_i\bm{\Psi}_i\bm{V}_i^H, \label{equ:SPrimeFormula}
\end{equation}
where $\bm{\Psi}_i$ is a $r_i\times r_i$ diagonal matrix with diagonal elements $\psi_{i_j} = \max (0,1-\frac{\delta_i}{\theta_{i_j}^2}),\ \forall j \in \{ 1,...,r_i \}$ ($\theta_{i_j}$'s are the diagonal elements of $\bm{\Theta}_i$).

The obtained optimal values for $\xi$, $\gamma_i$'s, $\bm{S}'_{AP}$, and $\bm{S}'_i$ are put into the problem \eqref{equ:DualMaxMinInfo}, and hence, the following optimization problem can be written for the remaining parameters:
\begin{maxi!}
{\bm{\delta},\zeta}{\min_{\bm{\tau}}\ \tau_u^i r_i(G_i+Z_i)\bigg[\log{(\delta_i)}-\frac{\delta_i}{r_i}\sum_{j=1}^{r_i}\frac{1}{|\theta_{i_j}|^2}\nonumber}{\label{equ:DualMaxMin}}{}
\breakObjective{+1-\frac{1}{r_i}\sum_{j=1}^{r_i}\log{|\theta_{i_j}|^2}\bigg]+\zeta(\tau_u^i -1) \nonumber}
\breakObjective{+(\frac{\zeta}{P_{peak}}+Z_{AP}-\lambda_{1,B})\tau_0 P_{peak} }
\addConstraint{\delta_i,\zeta \ge 0,\ \forall i \in \{ 1,...,K \}}.
\end{maxi!}
It should be noted that as E-R$_i$ receives almost all the power (the E-AP transfers power toward this E-R), $\tau_u^j$ will be zero $\forall j\ne i$. The above problem is a linear function of $\tau_0$ and $\tau_u^i$. In order to have a nonzero throughput, we must have $\tau_0,\tau_u^i>0$, and 
as a result, the slopes of $\tau_0$ and $\tau_u^i$ must be zero in the optimal solution. Therefore, the following equations can be written:
\begin{align}
-\log{\delta_i}&+\frac{1}{r_i}\sum_{j=1}^{r_i}\log{|\theta_{i_j}|^2}-1+\frac{\delta_i}{r_i}\sum_{j=1}^{r_i} \frac{1}{|\theta_{i_j}|^2}\nonumber\\&-\frac{\zeta}{r_i(G_i+Z_i)} = 0,  \label{equ:DeltaZeta} \\
\lambda_{\bm{B},1} &- Z_{AP}-\frac{\zeta}{P_{peak}}=0, \label{equ:DeltaZetaAP}
\end{align}
where $\lambda_{\bm{B},1} = \delta_i\lambda_{\bm{W}_i,1}$ and $\lambda_{\bm{W}_i,1}$ is the maximum eigenvalue of $\bm{W}_i$. Using equations \eqref{equ:DeltaZeta} and \eqref{equ:DeltaZetaAP}, the optimal $\delta_i$  can be written   as follows:
\begin{equation}
\delta_i = \frac{1}{\beta_i}\mathcal{W}(\beta_ie^{\alpha_i}),
\end{equation}
 where $\mathcal{W}(.)$ is the Lambert W function and 
 \begin{align}
 \alpha_i &\triangleq \frac{1}{r_i}\sum_{j=1}^{r_i}\log{|\theta_{i_j}|^2}+\frac{Z_{AP}P_{peak}}{r_i(G_i+Z_i)}-1,\\ \beta_i &\triangleq \frac{1}{r_i}(\frac{\lambda_{\bm{W}_i,1}P_{peak}}{G_i+Z_i}-\sum_{j=1}^{r_i}\frac{1}{|\theta_{i_j}|^2}).
\end{align} 
The Lambert W function has real values for $\beta_ie^{\alpha_i}\ge -e^{-1}$, which means that if this inequality does not hold for the obtained optimal values, then the objective function value for E-R$_i$ will be zero. 
Without loss of generality, we assume that  E-R$_i$ has the minimum objective value among all the E-Rs. Then, $\tau_0$ and $\tau_u^i$ can be obtained using the related KKT conditions of {\color{black}problem formulation \eqref{equ:DualMaxMinInfo}} as follows:
\begin{align}
Tr(\bm{S}'_i) &= Tr(\bm{W}_i\bm{S}'_{AP}),  \label{equ:Tau0Taui} \\
\tau_0+{\tau_u^i} &= 1. \label{equ:someTau}
\end{align}
Using equations \eqref{equ:SBSPrimeFormula} and \eqref{equ:SPrimeFormula} for $\bm{S}'_{AP}$ and $\bm{S}'_i$, the optimal $\tau_0$ and $\tau_u^i$ can be obtained as follows:
\begin{equation} 
\tau_0 = \frac{1}{1+ \omega_i}, \ \tau_u^{i} = \omega_i \tau_0, 
\end{equation}
where $\omega_i \triangleq \frac{\delta_i P_{peak}Tr(\bm{W}_i\bm{u}_{\bm{B},1}\bm{u}_{\bm{B},1}^H)}{(G_i+Z_i)\sum_{j=1}^{r_i}\psi_{i_j}}$. This completes the proof of Theorem \ref{theo:QPF-IT}.




\balance



%
%
%

\bibliographystyle{IEEEtran}
\bibliography{WPT_Globecom_references}

\end{document}